\documentclass{jfm}
\usepackage{graphicx}
\usepackage{epstopdf, epsfig}

\usepackage{subcaption} 
\usepackage{xcolor} 
\usepackage{amsmath}
\usepackage{epstopdf} 
\usepackage[retainorgcmds]{IEEEtrantools}
\usepackage[colorlinks=true,linkcolor=red,citecolor=blue]{hyperref}

\newcommand{\ud}{\,\mathrm{d}}
\newcommand{\e}{\mathrm{e}}

\newcommand{\SI}[1]{\:\mathrm{#1}}

\renewcommand{\i}{\mathrm{i}}
\newcommand{\add}[1]{{\color{black}#1}}

\graphicspath{{Figs/}{IllustrationFigs/}{DataFigs/}}
\shorttitle{Canonical correlation decomposition}
\shortauthor{B. Lyu}

\title{Canonical correlation decomposition of numerical and experimental data
for observable diagnosis} 

\author{B. Lyu\aff{1}
    \corresp{\email{b.lyu@pku.edu.cn}} 
}

\affiliation{\aff{1}State Key Laboratory for Turbulence and Complex Systems,
College of Engineering, Peking University, Beijing 100871, China}

\begin{document}
\maketitle 
\begin{abstract}     
A flow decomposition method based on canonical correlation analysis is proposed
in this paper to optimally dissect complex flows into mutually orthogonal modes
that are ranked by their cross-correlation with an observable. It is
particularly suitable for identifying the observable-correlated flow structures
while effectively excluding those uncorrelated, even though they may be highly
energetic. Therefore, this method is capable of extracting coherent flow
features under low signal-to-noise ratios. A numerical validation is conducted
and shows that the method can robustly identify the observable-correlated flow
events even though the underlying signal is corrupted by random noise that is
four orders of magnitude stronger. The temporal sampling frequency and duration
of the observable determine the maximum and minimum frequencies to be resolved
in the cross-correlation respectively, while those of the flow are to ensure
convergence. These criteria are validated using synthetic examples. The
decomposition method is subsequently used to analyse a turbulent channel flow, a
subsonic turbulent jet and an unsteady vortex shedding from a cylinder, showing
the effectiveness of observable-correlated structure identification and order
reduction. This decomposition represents a data-driven method of effective order
reduction for highly noisy numerical and experimental data and is suitable for
identifying the source and descendent events of a given observable. It is hoped
that this method will join the existing flow diagnosis tools, in particular for
observable-related diagnosis and control. 
\end{abstract}

\section{Introduction}
\label{sec:intro}
Many natural flows exhibit complex behaviour, such as the boundary layer formed
over a sand dune or compressed air inside an aeroengine. This is particularly
true at high Reynolds numbers, where most realistic engineering flows occur,
because turbulence comes into play exhibiting a wide range of temporal and
spatial scales. To understand, model, and possibly exert control on these flows,
it is crucial to extract dominant structures and reduce the systems' degrees of
freedom.

Extensive research has been conducted to extract coherent features and decompose
complex flows into a collection of simple modes. Well-established methods
include the Proper Orthogonal Decomposition (POD), Dynamic Mode Decomposition
(DMD), \add{resolvent/input-output analysis and global stability
analysis~\citep{Taira2017}. Among these, POD and DMD fall into the category of
data-driven approaches, while the resolvent/input-output analysis and global
stability analysis are model-based.}

POD~\citep{Lumley1967,Berkooz1993} is a particularly well-known data-driven
method and represents a powerful tool for feature extraction and order
reduction. Originating from Principal Component Analysis (PCA) in classical
statistics, POD decomposes a complex flow into mutually orthogonal modes ranked
by their fluctuation energy. If a flow is comprised of a few energetic coherent
structures, POD effectively identifies them as leading-order modes. A linear
combination of these leading-order modes then forms an optimal reduced-order
representation of the total flow. POD may be used to extract the spatial or
temporal structures~\citep{Lumley1970, Sirovich1987}. These two structures are
coupled, with the temporal structures representing the temporal variation of
their corresponding spatial modes, and the spatial modes representing the
spatial distribution of their corresponding temporal modes~\citep{Aubry1991}.
This leads to the so-called Bi-orthogonal Decomposition (BOD). Recent years have
also seen the increasingly widely-used Spectral Proper Orthogonal Decomposition
(SPOD) in studying turbulent flows~\citep{Towne2018}. \add{In addition, to
better capture the structures in transient and intermittent flows, conditional
space-time POD~\citep{Schmidt2019} and multidimensional empirical mode
decompositions~\citep{Souza2024} are proposed. These techniques are used to
examine the acoustic bursts, the onset and evolution of the dynamic stall and
intermittent vortex pairs, showing advantageous capability in resolving
transient and intermittent events. It is worth noting that since POD relies on
the underlying coherence within the flow to work, it is capable of identifying
the flow structures that are dynamically nonlinear compared to linear
model-based approaches.}

While POD aims to identify the coherent structures within a complex flow, DMD
aims to extract temporal evolutionary information of the underlying dynamics
captured in the data~\citep{Schmid2010}. The resulting representation is a
dynamical system of fewer degrees of freedom. DMD starts by assuming a linear
mapping between a sequence of the flow data, and the dynamics is extracted by
examining the eigenvalues of a similarity matrix. For a linear system, this
amounts to identifying the eigenmodes of the system. For nonlinear systems, DMD
is connected with the modes of the so-called Koopman
operator~\citep{Koopman1931,Mezic2013,Schmid2022}. Unlike POD, \add{DMD modes
capture the main ``contributions'' to the overall dynamics embedded in the data
sequence}. Recent years have seen numerous variants of DMD such as the extended
DMD~\citep{Williams2015} and Residual DMD~\citep{Colbrook2023}. More details on
the recent development of DMD can be found in the recent review by
\citet{Schmid2022}.

As mentioned above, both POD and DMD are data-driven, while the resolvent
analysis is based on the modal analysis of a linear operator. The resolvent
analysis has an early origin in control theory and is based on the
pseudospectrum of an operator~\citep{Trefethen1993,Taira2017}, rather than the
spectrum. \add{For example, when the flow is decomposed into a base part and a
fluctuation part, the Navier-Stokes equations can be rewritten and interpreted
as a forced linear system, by which the evolution of the fluctuation part is
governed.} The nonlinear terms are collected on the right-hand side and
interpreted as the forcing of the system. The resolvent modes are ranked by the
energy gain between the response and forcing. Therefore, the resolvent analysis
examines the gain properties of the linearized operator and has been
successfully used to study turbulence from a linearized Navier-Stokes equation
point of view~\citep{Farrell1993,Mckeon2010}. Recent studies also show that the
leading-order resolvent modes match the leading-order \add{SPOD} modes extracted
from a numerically simulated high-speed jet~\citep{Schmidt2018}. The
input-output analysis~\citep{Jovanovic2021} is similar to the resolvent analysis
in that a modal analysis is performed on a linearized operator. Input-output
analysis differs from the conventional resolvent analysis in that a weight may
be added to the operator to bias both the forcing and response towards
interested domains or observables~\citep{Jeun2016}. Therefore, input-output
analysis may be regarded as a weighted resolvent analysis. 

\add{In contrast to the resolvent analysis, model-driven global stability
analysis~\citep{Theofilis2011} examines the eigenvalue properties of an operator
linearized around a base flow with multiple inhomogeneous spatial directions. In
particular, it pays special attention to unstable modes, which would dominate
the linear response of the system at large times. Note that through global
stability analysis, the stable modes can also be obtained, which may play an
important role in determining the transient dynamics of underlying flows. This
is particularly true in fluid mechanics, where the linearized operators are
often non-normal~\citep{Trefethen1993} and the transient growth can become
crucial in determining the flow stability. In addition, an adjoint analysis of
the operator may be performed to examine the receptivity problem, yielding modes
that are similar to the optimal forcing modes in the resolvent analysis.} 

POD and DMD, together with their variants, are common data-driven flow
decomposition methods used in fluid mechanics. These provide important tools for
probing the structures and dynamics of an underlying dynamical system. The
ultimate goal of identifying the dominant structures or dynamics is, however,
often to understand and possibly control some observables of the flow, such as
to reduce the drag of a cylinder, minimise the unsteady force of a wing, or
abate the noise emission from a jet. However, because POD modes are ranked by
their fluctuation energy, the leading-order modes are not necessarily the most
important structures as far as the observable is concerned, although they do
carry the largest energy. For example, a large coherent structure effectively
extracted from a turbulent subsonic jet using POD may be very inefficient at
generating noise. In other words, the leading-order POD mode may not be the
leading-order noise-generating flow structure. For example, it has been shown
that a substantial number of near-field POD modes are required to
reconstruct the acoustic field~\citep{Freund2009}. Similarly, DMD extracts
the dominant dynamics embedded within the flow without taking their connection
with any observable into account. Consequently, the leading-order dynamic mode
does not necessarily represent the flow events connected with the leading-order
dynamics of the observables. 

\add{That the energy rank may not be an appropriate measure, in particular for
an observable-related diagnosis, is a well-recognised limitation of
POD~\citep{Rowley2005,Schmid2010}}. One widely-used approach to overcome this
difficulty is to use different norms to bias the decomposition towards
interested observables or to use the extended POD~\citep{Maurel2001,Boree2003}.
For example, \citet{Freund2009} performed the POD decomposition of a turbulent
jet using various norms, including the near-field turbulent kinetic energy,
near-field pressure, and far-field pressure. When the far-field pressure norm is
used, the near-field flow quantities drop out in the correlation matrix and the
resulting modes are effectively ranked only by the far-field pressure. Although
the near-field flow can still be projected onto the far-field basis, the
resulting \add{near-field} mode does not necessarily form a direct continuation
of the far-field physics, particularly when the near- and far-field exhibit
completely different dynamics or the far-field and near-field variables are
characterised by pronounced phase delays. \add{Note that the balanced POD
proposed by \citet{Rowley2005} is another similar technique to overcome the
energy norm limitation of POD, which may be viewed as a special form of POD when
the observability Gramian is used as the norm.}

On the other hand, the resolvent and input-output analyses decompose the flow to
maximise the energy gain between the output and forcing based on the spectral
theory of linear operators. \add{Hence, the observable may be directly included
in the choice of output.} The resolvent and input-output analysis represent
powerful tools to diagnose the flow structure and are capable of providing
insightful understanding into a variety of turbulent
flows~\citep{Mckeon2010,Sharma2013}. In order to do so, a linearized operator
describing the underlying system is often needed. In some cases, however, such
an operator may not be readily known, while in others the linearized operator
may not be an appropriate representation of the dynamical system, particularly
in highly nonlinear systems. For example, \add{an input-output analysis} was
performed on compressible subsonic and supersonic jets and found that a
considerable number of modes were required to reconstruct the acoustic energy of
subsonic jets~\citep{Jeun2016}, which may be partly due to the limitation
imposed by linearity. \add{Such a limitation is also applicable to global
stability analysis, where a linearized operator must be known in advance.}


In this paper, we aim to develop a data-driven flow decomposition method that is
suitable for observable diagnosis based on flow and observable snapshots instead
of linear operators. Instead of redefining the POD energy norm to bias towards
the observable, the decomposition aims to introduce a rank based on a
cross-correlation norm between the resulting modes and the observable, hence
including both the flow and observable data in the correlation matrix. \add{The
decomposition method falls under the framework of canonical correlation analysis
(CCA)~\citep{Hotelling1936} in classical statistics.}
\add{This paper is structured as follows: section~\ref{sec:ccd} shows a
mathematical formulation of the decomposition method. The physical significance
of the resulting modes, the frequency and wavenumber resolutions, the effect of
including multiple observables and the connection of the present decomposition
to POD and the extended POD are discussed in detail sequentially.
Section~\ref{sec:validation} validates the method by performing the
decomposition on multiple synthetic flow fields. The effects of varying sampling
frequency, duration and including multiple observables are also thoroughly
validated.} Section~\ref{sec:realFlow} applies this technique to both numerical
and experimental data, demonstrating the potential use of such a method. The
following section concludes the paper and lists some future work. 

\section{The canonical correlation decomposition} 
\label{sec:ccd}

\subsection{The decomposition procedure}
\label{subsec:ccd}
\begin{figure}
    \centering
    \includegraphics[width=0.35\textwidth]{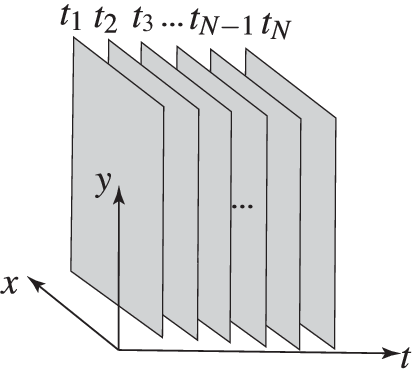}
    \caption{Schematic illustration of the two-dimensional flow snapshots
    sampled at time $t_i$, $i=1,2,3,\ldots, N$. Each snapshot contains flow data
    in both $x$ and $y$ directions, where $x$ and $y$ denotes the Cartesian
    coordinates of the flow domain. }
    \label{fig:FlowSnapshots}
\end{figure}

Assume that we have a sequence of snapshots $\boldsymbol{u}_i$ obtained by
sampling a flow field $u(\boldsymbol{x}, t)$ at time $t = t_i$, where
$\boldsymbol{x}$ represents the coordinates of the flow domain and $i$ is an
integer that takes the value of $1,2,3,\ldots, N$. If the snapshots
$\boldsymbol{u}_i$ are sampled in time, $t_{i}$ increases sequentially  as $i$
increases, as shown in figure~\ref{fig:FlowSnapshots}. If $\boldsymbol{u}_i$
are, however, sampled in the ensemble space, each $t_i$ refers to the sampling
time in its corresponding independent realisation and can therefore be
independent of each other. \add{In the most general case, $\boldsymbol{u}_i$ can
be sampled both in the time and ensemble space.} Each snapshot of this sequence
is obtained by discretizing the spatial domain on a mesh and represented by a
column vector of length $M$. We write this snapshot sequence compactly in a
matrix notation as
\begin{equation}
    \boldsymbol{U} = \left[\boldsymbol{u}_1, \boldsymbol{u}_2,
    \boldsymbol{u}_3,\ldots, \boldsymbol{u}_N\right].
    \label{equ:Umatrix}
\end{equation}

For each snapshot $\boldsymbol{u}_i$, which is obtained by sampling a flow field
at $t_i$, assume that we can simultaneously sample an interested observable
$p(t)$ of this flow at time $t_i+\tau_j$, where $j$ is an integer and takes the
values of $1,2,3,...Q$ with $Q$ being a positive integer. We therefore obtain a
sequence of the sampled observable $p_{i, j}$, $j=1,2,3...Q$. Note that the
sequence $p_{i,j}$ can be sampled at an earlier or later time of $t_{i}$,
depending on whether $\tau_1$ is a negative or positive value, respectively.
This is important, and we will discuss its significance in the rest of this
paper.

For each integer $i$, we can define a column vector $\boldsymbol{p}_i$ such that
\begin{equation}
    \boldsymbol{p}_i = [p_{i,1}, p_{i,2}, p_{i,3}, \ldots p_{i,Q}]^T,
    \label{equ:pvector}
\end{equation}
where $T$ denotes transpose.
We then form a matrix $\boldsymbol{P}$ such that
\begin{equation}
    \boldsymbol{P}=[\boldsymbol{p}_{1}, \boldsymbol{p}_2, \ldots,
    \boldsymbol{p}_{N-1}, \boldsymbol{p}_N].
    \label{equ:PMatrix}
\end{equation}
The key step is to construct a matrix $\boldsymbol{A}$, representing the
cross-correlation matrix between the flow and the observable, such that
\begin{equation}
    \boldsymbol{A} = \frac{1}{N\sqrt{Q}} \boldsymbol{P} \boldsymbol{U}^\dagger,
    \label{equ:Amatrix}
\end{equation}
where $\dagger$ denotes the \add{Hermitian adjoint}. The \add{Hermitian adjoint}
here allows both $\boldsymbol{P}$ and $\boldsymbol{U}$ to be complex matrices.
This is useful because both the observable and flow field can be just a Fourier
component of the total fields \add{(see the end of section~\ref{subsec:ccd} for
more details)}. In the case where only real matrices are involved, the
\add{Hermitian adjoint} $\dagger$ reduces to the simple transpose $T$. 

We then perform the standard Singular Value Decomposition (SVD) of matrix
$\boldsymbol{A}$, such that
\begin{equation}
    \boldsymbol{A} = \boldsymbol{R} \boldsymbol{\Sigma} \boldsymbol{V}^\dagger,
    \label{equ:svd}
\end{equation}
where $\boldsymbol{R}$ and $\boldsymbol{V}$ are $Q \times Q$ and $M \times M$
unitary matrices respectively, while $\boldsymbol{\Sigma}$ is a diagonal matrix
of $Q \times M$ with the singular values $\sigma_j$ ($j=1, 2, 3\ldots, \min(M,
Q)$) as its diagonal elements. The column vectors of $\boldsymbol{V}$ represent
the desired modes of the flow field $\boldsymbol{u}_i$, while those of
$\boldsymbol{R}$ represent the normalised cross-correlation functions between
the resulting modes and the observable. From SVD, it can be readily shown that
these modes are mutually orthonormal and form a complete basis of
$\mathbb{R}^M$. Therefore, the flow field $\boldsymbol{u}_i$ can be conveniently
decomposed as 
\begin{equation}
    \boldsymbol{u}_i = \sum_{k = 1}^N a_k(t_i) \boldsymbol{v}_k,
    \label{equ:expansionDiscretized}
\end{equation}
where $\boldsymbol{v}_k$ denotes the $k$-th column of $\boldsymbol{V}$ while
$a_k(t_i)$ denotes its corresponding expansion coefficient at time $t_i$, or
equivalently,
\begin{equation}
    u(\boldsymbol{x}, t) = \sum_{k = 1}^N a_k(t) \phi_k(\boldsymbol{x}),
    \label{equ:expansionContinous}
\end{equation}
where $\phi_k(\boldsymbol{x})$ denotes the basis function corresponding to
$\boldsymbol{v}_k$, while $a_k(t)$ is the expansion coefficient of $u(
\boldsymbol{x}, t)$ using the basis $\phi_k(\boldsymbol{x})$. As shown in
section~\ref{subsec:physicialSignificance}, these modes are ranked by their
cross-correlation with the observable, and a significant order reduction may be
expected if only a small number of modes are pronouncedly correlated with the
observable. \add{As will be shown in section~\ref{subsec:physicialSignificance},
the decomposition method falls under the framework of CCA, therefore it will be
referred to as the canonical correlation decomposition (CCD) in the rest of this
paper.}

\add{Note that, as mentioned above, both the observable and the flow can be just
a Fourier component of the total fields. For example, the observable may be
$\tilde{p}_\omega$ while the flow may be $\tilde{\boldsymbol{u}}_\omega$, where
$\tilde{p}_\omega$ and $\tilde{\boldsymbol{u}}_\omega$ represent the temporal
Fourier components of the observable and flow at angular the frequency $\omega$,
respectively. In practice, a long flow snapshot sequence $u(\boldsymbol{x},
t_k)$ ($k=1,2,3\ldots$) obtained in experiments or simulations may be first
partitioned into $N$ segments; each segment may be regarded as a realisation in
the ensemble space and then Fourier transformed in time and/or space to form the
$\boldsymbol{u}_i$ ($i=1,2\ldots N)$ shown in (\ref{equ:Umatrix}). Similarly, a
long observable sequence $p(t_k+\tau_1)$ obtained in experiments or simulations
may be partitioned into $N$ segments; the $i$th segment is then Fourier
transformed with respect to $t_k$ ($\tau_1$ is a constant) to obtain
$\tilde{p}_{\omega i}(\tau_1)$. In a similar manner, $\tilde{p}_{\omega
i}(\tau_2)$, $\tilde{p}_{\omega i}(\tau_3),\ldots$ $\tilde{p}_{\omega
i}(\tau_Q)$ can be obtained, which are just $p_{i,2}$, $p_{i,3},\ldots$
$p_{i,Q}$ shown in (\ref{equ:pvector}) ($\tilde{p}_{\omega i}(\tau_1)$
constitutes $p_{i,1}$). Note that when the observable and the flow are sampled
at different frequencies, proper temporal alignment of them for each realisation
must be ensured according to those described in section~\ref{subsec:ccd}. Care
must also be taken regarding the frequency resolutions of the flow and the
observable when the Fourier transform is performed. CCD can then be performed
according to (\ref{equ:PMatrix}) to (\ref{equ:svd}), which may be regarded as a
form of CCD decomposition in the spectral space. }

\subsection{Physical significance of CCD modes}
\label{subsec:physicialSignificance}
The CCD represents an optimal decomposition that maximises the cross-correlation
between the flow field $u$ and the observable $p$. This can be shown
mathematically as follows. Assuming the flow field is described by the function
$u(\boldsymbol{x}, t)$ while the observable by $p(t+\tau)$, where $\tau$
represents the time delay between flow and the observable. We form the
cross-correlation $R(\tau, \boldsymbol{x})$ using
\begin{equation}
    R(\tau, \boldsymbol{x}) = \left\langle{p^\ast(t+\tau)
    u(\boldsymbol{x}, t)}\right\rangle,
    \label{equ:correlationR}
\end{equation}
\add{where $*$ represent the complex conjugate, while $\langle \cdot \rangle$
represents the temporal or ensemble average. In the latter case, the statistical
processes represented by $u$ and $p$ are assumed to be stationary. For
non-stationary processes, (\ref{equ:correlationR}) explicitly depends on $t$,
but the following derivation can still proceed. }

First, let us define an inner product in the Hilbert space defined on a domain
$\Omega$ such that
\begin{equation}
    (f, g) = \int_\Omega f(\boldsymbol{x}) g^\ast(\boldsymbol{x}) \ud x^n,
\end{equation}
where $f(\boldsymbol{x})$ and $g(\boldsymbol{x})$ denote two functions within
this space and $n$ represents the dimension of $\Omega$. A norm is therefore
defined as $||f||= (f, f)^{1/2}$. Suppose we wish to find a function
$\phi(\boldsymbol{x})$ of unit norm, such that the inner product between
$R(\tau, \boldsymbol{x})$ and $\phi(\boldsymbol{x})$, i.e. $(R, \phi)$, obtains
its maximum value in the $L_2$ norm.  \add{Mathematically, this is equivalent
to}
\begin{equation}
    \max_{||\phi||=1} \frac{1}{T} \int_{\tau_0}^{\tau_0 + T}|(R,
    \phi)|^2 \ud \tau,
    \label{equ:RDefinition}
\end{equation}
\add{where $|\cdot|$ represents the complex modulus, and $\tau_0$ and $T$} are
two constants chosen such that the integration includes the entire interval
where the integrand obtains non-negligible values.

Physically, this amounts to finding the optimal function $\phi$ that most
correlates with the observable. This is because the ensemble average in
(\ref{equ:correlationR}) commutes with the inner product in
(\ref{equ:RDefinition}), i.e. 
\begin{equation}
    (R, \phi) = \langle p^\ast(t+\tau)a_\phi(t) \rangle,
    \label{equ:RExplanation}
\end{equation}
where $a_\phi(t)$ represents the expansion coefficient of the flow field $u$
using the basis $\phi$, i.e.
\begin{equation}
    a_\phi(t) = (u, \phi).
    \label{equ:akDef}
\end{equation}
Clearly, we see from (\ref{equ:RExplanation}) and (\ref{equ:akDef}) that $(R,
\phi)$ represents the cross-correlation function between the mode $\phi$ and the
observable $p$. The $L_2$ norm of $(R, \phi)$ defined over an interval of length
$T$ is a natural measure of the correlation level between $\phi$ and $p$.  We
therefore define the correlation strength $C_e$ as the average of $ |(R,
\phi)|^2 $ over the interval $[\tau_0, \tau_0+T]$, i.e. 
\begin{equation}
    C_e = \frac{1}{T}\int_{\tau_0}^{\tau_0+T} |(R, \phi)|^2 \ud \tau.
    \label{equ:CeDef}
\end{equation}
Evidently, if
$\phi(\boldsymbol{x})$ maximises $C_e$, it represents a flow structure that most
correlates with the observable $p$. 

The function $\phi(\boldsymbol{x})$ that we seek can be obtained from an
eigenvalue problem as follows. We know that $\phi(\boldsymbol{x})$ is \add{a
function of unit norm that yields a maximum $C_e$, i.e. $\phi(x)$ satisfies}
\begin{equation}
    \max_{||\phi||=1} \frac{1}{T} \int_{\tau_0}^{\tau_0+T} |(R, \phi)|^2 \ud
    \tau.
    \label{equ:maxEquation}
\end{equation}
Classic calculus of variation shows that a necessary condition for
(\ref{equ:maxEquation}) to hold is that $\phi$ is an eigenfunction of the
correlation tensor, i.e. 
\begin{equation}
    \int_{\Omega} B(\boldsymbol{x}, \boldsymbol{x}^\prime)
    \phi(\boldsymbol{x}^\prime) \ud x^{\prime n} = \lambda
    \phi(\boldsymbol{x}),
    \label{equ:BeigenValue}
\end{equation}
where the correlation tensor is defined by 
\begin{equation}
    B(\boldsymbol{x}, \boldsymbol{x}^\prime) =
    \frac{1}{T}\int_{\tau_0}^{\tau_0+T} 
    R(\tau, \boldsymbol{x})
    R^\ast(\tau, \boldsymbol{x}^\prime) \ud \tau,
    \label{equ:BDef}
\end{equation}
and the eigenvalue $\lambda$ corresponds to $C_e$ defined in
(\ref{equ:CeDef})~\citep{Riesz1955}. Clearly, the maximum $C_e$ is given by the
largest eigenvalue.

When the flow field and the observable are discretized, we can show that after
multiplied by $\sqrt{Q}$ the matrix $\boldsymbol{A}$ defined in
section~\ref{subsec:ccd} is identical to a discretized form of $R^\ast(\tau,
\boldsymbol{x}^\prime)$. The correlation tensor $B(\boldsymbol{x},
\boldsymbol{x}^\prime)$ then reduces to $\boldsymbol{A}^\dagger \boldsymbol{A}$
because
\begin{equation}
    B(\boldsymbol{x}, \boldsymbol{x}^\prime) =
    \frac{1}{T}\int_{\tau_0}^{\tau_0+T} R(\tau, \boldsymbol{x})
    R^\ast(\tau, \boldsymbol{x}^\prime) \ud \tau \approx
    \frac{1}{Q}\sum_{i=1}^Q R(\tau_i, \boldsymbol{x}) R^\ast(\tau_i,
    \boldsymbol{x}^\prime) = \boldsymbol{A}^\dagger \boldsymbol{A},
    \label{equ:AA}
\end{equation}   
where $\tau_i$ is the discretized values of $\tau$.
Equation~\ref{equ:BeigenValue} therefore reduces to a discretized eigenvalue problem of the
matrix $\boldsymbol{A}^\dagger \boldsymbol{A}$,  i.e. 
\begin{equation}
    \boldsymbol{A}^\dagger \boldsymbol{A} \boldsymbol{v}_k = \lambda_k
    \boldsymbol{v}_k,
    \label{equ:eigenValueProblem}
\end{equation}
where $\boldsymbol{v}_k$ as defined in section~\ref{subsec:ccd} is the
discretized form of the $k$-th eigenfunction $\phi(\boldsymbol{x})$, while
$\lambda_k$ is the $k$-th $\lambda$ in (\ref{equ:BeigenValue}) subject to a
discretization constant, \add{whose exact value often carries no significance in
practice}. The eigenvalue problem of (\ref{equ:eigenValueProblem}) is equivalent
to the singular value decomposition shown in (\ref{equ:svd}). Therefore, the
column vectors of $\boldsymbol{V}$ are these optimal modes, while the
corresponding column vectors of $\boldsymbol{R}$ are the normalised
cross-correlation functions. In addition, the squares of the singular values
$\sigma_k^2$ are precisely $\lambda_k$, representing the correlation strength
$C_e$ between the CCD modes and observable (subject to a discretization
constant). In particular, when the components of $p$ that correlate with their
corresponding CCD modes of $u$ are of equal energy, $\sigma_k^2$ also represent
the \textit{observable-correlated} energy of their corresponding CCD modes
(subject to a constant), and the correlation ranking is identical to the ranking
of the \textit{observable-correlated} flow energy. In summary, instead of
decomposing the flow field $u$ based on its energy ranking using the classical
POD, (\ref{equ:svd}) yields a decomposition that is based on a cross-correlation
ranking with an observable, or the observable-correlated energy ranking in the
special case where the correlated components of $p$ are of equal energy.

Mathematically, the flow decomposition method can be shown to fall under the
framework of CCA~\citep{Hotelling1936} as follows. Given two column vectors
$\boldsymbol{X}=(x_1, x_2,\ldots x_n)^T$ and $\boldsymbol{Y}=(y_1,y_2,\ldots
y_m)^T$ of random variables with finite second moments, CCA seeks two vectors
$\boldsymbol{a} (\boldsymbol{a} \in \mathbb{R}^n)$ and $\boldsymbol{b}
(\boldsymbol{b} \in \mathbb{R}^m)$ such that the random variables
$\boldsymbol{a}^T\boldsymbol{X}$ and $\boldsymbol{b}^T\boldsymbol{Y}$ yield the
maximum correlation. The process may be continued in a subspace to yield a
sequence of vector pairs. In the context of CCD, the flow field $u$ may be
regarded as the $\boldsymbol{Y}$ vector. However, the key part of the
decomposition is to find a proper $\boldsymbol{X}$ vector. There are many ways
$\boldsymbol{X}$ can be specified, such as the flow within a specific subdomain
of interest. However, the essence and novelty of the present decomposition is to
construct an $\boldsymbol{X}$ that consists of the observable sampled in a
synchronised manner with the flow but at different time delays. Compared to POD
or the extended POD, this time shift is an additional dimension used in CCD. As
will be shown, the additional information embedded in this ``hidden"
shifted-time dimension is the key to yielding a more observable-targeting
decomposition. More importantly, this permits independent sampling rates between
the observable and the flow, which can be of great advantage. 

CCD possesses a number of key features that would be particularly useful for
targeted flow diagnosis. First, the decomposition modes are not ranked by their
energy, but by the correlation strength with the observable. \add{Flow features
that are not correlated with the observable can be effectively suppressed, while
those correlated are promoted and ranked according to their correlation strength
with the observable.} This targets exclusively the observable and is, therefore,
very useful in finding the sources or descendent structures of the observable.
Second, as will be shown in section~\ref{sec:validation}, the decomposition is
robust even when the signal-to-noise ratio (SNR) is low. This is useful when
only a small portion of the flow energy correlates with the observable, for
example in the classical problem of aeroacoustic emission due to turbulence.
Moreover, this robustness can be continuously improved when a longer time
duration is used. This is therefore suitable for experimental diagnosis, where
an arbitrarily long measurement may be readily performed. 

Third, \add{as will be shown in section~\ref{sec:validation}}, the decomposition
appears more capable of order or dimensionality reduction compared to POD. This
is because CCD aims to decompose the flow only in the observable-correlated
subspace, rather than the entire $\mathbb{R}^M$. In fact, this fact may be used
to estimate the convergence of the decomposition by examining how well
observable can be reconstructed only using modes corresponding to non-zero
singular values. Last but not least, the flexibility to use different sampling
frequencies for the flow and the observable enables one to fully exploit the
instrument's capabilities in experiments and numerical simulations. For example,
it is well known that acoustic signals can often be sampled much faster using a
microphone than the entire flow field using PIV. Similarly, in numerical
simulations, the observable can also be sampled much faster than the flow due to
limited storage requirements imposed by the observable at only a number of probe
positions. \add{Note that in general the sampling frequency of the observable is
independent of that of the flow, provided $\boldsymbol{p}_i$ properly aligns
with $\boldsymbol{u}_i$ as prescribed in section~\ref{subsec:ccd}. In practice,
if the sampling frequency of the observable is an integer multiple of that of
the flow, it would be trivial to achieve such alignment. In other cases,
clock-triggered synchronisation may be used to meet such a requirement in
experiments.}


\subsection{Frequency and wavenumber resolution and sampling delay} 
\label{subsec:resolution}
In section~\ref{subsec:ccd} we mention that the matrix $\boldsymbol{P}$ is
assumed to have $Q$ rows and each adjacent row is shifted by time $\Delta \tau =
\tau_{j+1}-\tau_j$ (assuming a constant sampling frequency). Moreover, $p$ is
sampled temporally behind $u$ by a time $\tau_1$ (or ahead of $u$ if $\tau_1$ is
negative). In practice, the choice of $Q$, $\Delta \tau$, and $\tau_1$ has
significant physical implications.

First, we show that $\Delta \tau$ and $Q$ determine the maximum and minimum
frequencies that can be resolved in the cross-correlation between the flow and
the observable, respectively. To see this, we start by defining the correlation
tensor $C(\tau^\prime, \tau)$ as
\begin{equation}
    C(\tau, \tau^\prime) = \int_\Omega R^\ast(\tau, \boldsymbol{x})
    R(\tau^\prime, \boldsymbol{x}) \ud x^n.
    \label{equ:CDef}
\end{equation}
Similar to that shown in section~\ref{subsec:physicialSignificance}, we can show
that after discretization $C(\tau, \tau^\prime)$ reduces to $\boldsymbol{A}
\boldsymbol{A}^\dagger$ subject to a scaled constant. The eigenvalue $\lambda$
defined in (\ref{equ:BeigenValue}) can also be found by 
\begin{equation}
   \frac{1}{T}\int_{\tau_0}^{\tau_0+T} C(\tau, \tau^\prime)
    \psi(\tau^\prime) \ud \tau^\prime = \lambda
    \psi(\tau),
    \label{equ:CeigenValue}
\end{equation}
where $\psi(\tau)$ corresponds to the column vectors of matrix $\boldsymbol{R}$
defined in (\ref{equ:svd}) in a discretized form. When the function $C(\tau,
\tau^\prime)$ is of a homogeneous (stationary) form, i.e. 
\begin{equation}
    C(\tau, \tau^\prime) = C_0(\tau- \tau^\prime),
\end{equation}
equation (\ref{equ:CeigenValue}) reduces to a Fourier
expansion~\citep{Berkooz1993}, i.e. 
\begin{equation}
    \int_{\tau_0}^{\tau_0+T} C_0(\tau - \tau^\prime) \e^{\i 2 \pi f
    \tau^\prime} \ud \tau^\prime =  \lambda T \e^{\i 2 \pi f \tau},
    \label{equ:CFourierProblem}
\end{equation}
or equivalently, 
\begin{equation}
    C(\tau, \tau^\prime)
    = \sum_{n} \lambda_n T
    \e^{\i 2 \pi f_n (\tau - \tau^\prime)}.
    \label{equ:CFourierExpansion}
\end{equation}
Equation (\ref{equ:CFourierExpansion}) indicates that the function $C(\tau,
\tau^\prime)$ can be expanded into a Fourier series. When $C(\tau, \tau^\prime)$
is discretized, the well-known Nyquist's theorem demands that the sampling
frequency $f_s^p\equiv 1/\Delta \tau$ of the observable must be at least twice
as large as the highest frequency to be resolved. Similarly, the total sampling
duration $Q\Delta\tau$ determines the frequency resolution to be
$1/Q\Delta\tau$. When the function $C(\tau, \tau^\prime)$ is not a homogeneous
function, there are no general theorems, but we expect that the frequency
requirement remains similar to the homogeneous case. In summary, $\Delta \tau$
determines the maximum frequency  while $Q$ determines the frequency resolution
similar to those in the Discrete Fourier Transform (DFT). 

Second, we note that the choice of $\tau_1$ \add{depends on the physical time
delay between $p$ and $u$}. In general, the observable may be temporally ahead
of or behind the flow events depending on the causal relations between the two.
For example, if $p$ represents the upstream forcing imposed near the nozzle lip
of a turbulent jet, then there must exist a finite time delay between the
evolved downstream structure and $p$ due to the finite propagation speed of jet
instability waves. \add{In this case, $p$ is preferably sampled ahead of $u$ in
order to capture the physical correlation within a reasonably short sampling
duration of $p$. A good estimation of $\tau_1$ would be around $-d/U_c$, where
$d$ and $U_c$ represent the maximum distance between the flow and the observable
and the convection velocity of the instability waves, respectively.} Conversely,
if $p$ is temporally behind $u$ then it must be sampled after $u$. \add{For
example, if the observable $p$ represents the acoustic pressure at a distance
$r$ from the jet flow, a good estimate of $\tau_1$ would be around $r/c$, where
$c$ represents the speed of sound. In other more general flows, a good estimate
of $\tau_1$ may be obtained by examining the cross-correlation function between
the flow and the observable. $\tau_1$ should be chosen such that the correlation
matrix $\boldsymbol{A}$ captures the entire correlation peaks.} 


In addition to the sampling rate, sampling duration, and sampling delay of $p$,
the temporal and spatial sampling of $u$ also have physical implications. First,
the spatial sampling rate of $u$ has the conventional implication that it
determines the maximum spatial wavenumber, whereas the length of the spatial
sample determines the minimum wavenumber that can be resolved. This can be shown
in a similar manner to those shown from (\ref{equ:CDef}) to
(\ref{equ:CFourierExpansion}) by considering the expansion of $B(
\boldsymbol{x}, \boldsymbol{x}^\prime)$. We omit a repetitive presentation here
for brevity. 

Second, the temporal sampling rate of the flow $f_s^u$ (when $\boldsymbol{u}_i$
is obtained via temporal sampling), however, has a different implication. The
sampling rate here is not to determine the frequency limit, but mainly to ensure
the convergence of the correlation between $p$ and $u$. In particular, there is
no need for the flow field and the observable to be sampled at the same
frequency. This is an important advantage, because, as discussed in
section~\ref{subsec:physicialSignificance}, in experiments PIV can only be
sampled at a much slower rate than that using a hot-wire or a microphone,
whereas in numerical simulations sampling the flow field fast is impractical
because of storage limit. However, such limitation does not exist for a number
of interested observables. Therefore, the much higher sampling rate of the
observable can be fully exploited by CCD in both experiments and numerical
simulations. The fact that the sampling rates of the flow and observable are
independent of each other is evident in the case that $\boldsymbol{u}_i$ is
obtained in the ensemble space.

\subsection{Inclusion of multiple observables}
\label{subsec:multiP}
In many applications, the appropriate observable is not necessarily limited by
one. For example, to examine the dominant flow structures in a subsonic round
jet that generates sound at $90^\circ$ to the jet centreline, the acoustic
pressure at any azimuthal position is an appropriate choice due to the azimuthal
statistical homogeneity. \add{In such cases, upon defining a local coordinate
system, each observable and the flow field in the local coordinates may be
treated as an independent realisation. In such cases, using multiple observables
is trivial by following section~\ref{subsec:ccd}, i.e. allowing
$\boldsymbol{u}_i $ to be sampled both in the temporal and ensemble space. By
doing so, the number of flow snapshots is increased by $N_{rl}$ fold, where
$N_{rl}$ denotes the number of independent realisations. This would be very
useful in improving the convergence of the resulting CCD modes.}

\add{In cases where there is no apparent statistical homogeneity in the flow,
multiple observables may still be included. For example, when a turbulent jet is
forced in an upstream position~\citep{Crow1971}, the introduced disturbance
evolves downstream. One may wish to extract the coherent structures induced by
the forcing using observable measurements downstream of the jet. In such cases,
velocity fluctuations at any location within a reasonable distance from the
forcing location may be used. However, each observable is likely to be heavily
contaminated by turbulence.} Using multiple observables are expected to improve
the converge of the resulting modes. In such case, suppose that the matrix
$\boldsymbol{P}_i$ ($i = 1, 2, 3\ldots L$) can be formed using the $i$-th
observable according to (\ref{equ:PMatrix}), then a straightforward way to
include multiple observables is to form the total matrix $\boldsymbol{P}$ such
that
\begin{equation}
    \boldsymbol{P} = \left[ 
	\begin{aligned}
	    \boldsymbol{P}_1\\ 
	    \boldsymbol{P}_2\\
	    \ldots\\
	    \boldsymbol{P}_L 
    \end{aligned}\right].
    \label{equ:multiPMatrix}
\end{equation}

\add{The normalisation constant $Q$ in (\ref{equ:Amatrix}) should be replaced by
$LQ$.} However, it is important to note that although $\boldsymbol{P}$ has $L$
times as many rows as $\boldsymbol{P}_i$, this does not improve the temporal
frequency resolution of the decomposition, which is still determined by
$\boldsymbol{P}_i$. This is because, as illustrated in
section~\ref{subsec:resolution}, the temporal frequency resolution is determined
by the duration of the time shift $Q\Delta\tau$ when (\ref{equ:correlationR}) is
truncated and discretized; including more observables does not increase the
length of this duration. Nevertheless, convergence of the resulting CCD modes
may improve due to the effective inclusion of more data, particularly when
highly noisy observables are used. For highly complicated flow data with a
limited sampling duration, such as those obtained in numerical turbulent
simulations, including multiple observables is expected to improve the
convergence, i.e. reduce the uncertainty or noise of the resulting CCD modes.

The choice of multiple observables, non matter in statistical homogeneous or
inhomogeneous flows, must be made with care. As mentioned, the observables must
be expected to resolve the same structures either due to statistical homogeneity
or well-defined sources of the underlying problem. In the case where the
multiple observables chosen are correlated with different events, including more
observables would effectively seek an average between these flow structures,
which may not be one's intention. For example, if one is interested in
identifying the flow structures that are most correlated with the skin friction
under a turbulent boundary layer, observables sampled at various streamwise
stations are expected to resolve different structures. In such cases, using
multiple observables may not be a worthwhile technique. 


\subsection{Connection to POD and extended POD} 
\label{subsec:connectionPOD}
As shown in section~\ref{subsec:physicialSignificance}, CCD is different from
POD in that the decomposition is based on a cross-correlation rather than an
energy norm. This difference is similar to that between CCA and its sister
method PCA in classical statistics. Physically, CCA aims to find the ``common
parts'' between two sets of variables, while PCA aims to find the main energetic
structures. Mathematically, instead of decomposing the matrix
$\boldsymbol{U}^\dagger$, a projection onto $\boldsymbol{P}$ is performed first
in CCD. This shows that the decomposition takes into account the space spanned
by $\boldsymbol{P}$. Note this projection may result in a rank that is lower
than that of the original flow; however, this is intended as one seeks to
decompose $\boldsymbol{U}^\dagger$ in the subspace correlated with the
observable only. One could argue that this projection leads to a ``lower-rank''
behaviour by construction, \add{as this would yield fewer singular values.
However, the low-rank behaviour we discuss in the following sections is not
characterised by fewer singular values, but rather characterised by a quick
decay of singular values as the mode number increases and, perhaps
more importantly, by a rapid reconstruction of the observable using fewer flow
modes.}

\add{As mentioned in section~\ref{sec:intro}, the extended
POD~\citep{Maurel2001,Boree2003} is developed with a similar aim as the present
decomposition, i.e. to better target the observable. One can show that the
extended POD using a subdomain $\boldsymbol{s}$ is closely related to the
degenerate case of CCD when no time shift is allowed between the observable and
flow (using multiple observables in $\boldsymbol{s}$). Mathematically, this
implies $Q=1$, $\tau_1=0$ and the observable matrix $\boldsymbol{P}$ shown in
(\ref{equ:PMatrix}) is a degenerate row vector of rank $1$. In the special case
where the subdomain of the extended POD only includes one observable point and
only one mode results, the extended POD and degenerate CCD are identical subject
to a normalisation constant. This can be shown as follows.

Suppose that there exist $L$ observables in the subdomain $\boldsymbol{s}$.
Since no time shift is allowed between the flow and observable, the matrix
$\boldsymbol{P}_i$ for each observable is a row vector. Hence, the assembled
matrix $\boldsymbol{P}$ is a matrix of dimension $L \times N$. Written in the
matrix convention used in the present paper, the essential steps of the spatial
extended POD start by decomposing $\boldsymbol{P}$ using POD or, equivalently,
by SVD, i.e.
\begin{equation}
    \boldsymbol{P}^\dagger = \boldsymbol{R}_s \boldsymbol{\Sigma}_s
    \boldsymbol{V}_s^\dagger,
    \label{equ:subdomainPOD}
\end{equation}
where both $\boldsymbol{R}_s$ and $\boldsymbol{V}_s$ are unitary matrices, the
subscript $s$ represents that this is a POD performed in the subdomain
$\boldsymbol{s}$. Note that it is the $\boldsymbol{P}^\dagger$ that is
decomposed. Right-multiplying (\ref{equ:subdomainPOD}) by $\boldsymbol{V}_s$,
one obtains 
\begin{equation}
    \boldsymbol{P}^\dagger \boldsymbol{V}_s = \boldsymbol{R}_s
    \boldsymbol{\Sigma}_s.
    \label{equ:matrixIntermedidate}
\end{equation}

Taking the $k$th column of both sides of (\ref{equ:matrixIntermedidate}) yields
\begin{equation}
    \boldsymbol{P}^\dagger \boldsymbol{V}_{s,k} = \boldsymbol{R}_{s,k}
    \sigma_{s,k},
    \label{equ:kele}
\end{equation}
where $\sigma_{s,k}$ represents the $k$th diagonal element of
$\boldsymbol{\Sigma}_s$. The right-hand side of (\ref{equ:kele}) represents the
temporal coefficient of the $k$th subdomain POD mode $\boldsymbol{V}_{s,k}$. The
$k$th extended POD mode $\boldsymbol{V}_{e,k}$ is obtained by projecting the
flow $\boldsymbol{U}$ in the extended domain defined in (\ref{equ:Umatrix}) onto
the $k$th temporal coefficient, followed by a normalisation, i.e. 
\begin{equation}
    \boldsymbol{V}_{e,k} = \frac{1}{\sigma_{s,k}^2}\boldsymbol{U}
    \boldsymbol{P}^\dagger \boldsymbol{V}_{s,k}.
    \label{equ:kthePOD}
\end{equation}

Following the procedure introduced in sections~\ref{subsec:ccd} and
\ref{subsec:multiP}, the multiple-observable CCD yields
\begin{equation}
    \frac{1}{N\sqrt{L}} \boldsymbol{P} \boldsymbol{U}^\dagger = \boldsymbol{R}
    \boldsymbol{\Sigma} \boldsymbol{V}^\dagger,  
    \label{equ:multiPProof}
\end{equation}
where $\boldsymbol{R}$, $\boldsymbol{\Sigma}$ and $\boldsymbol{V}$ are defined
earlier in section~\ref{subsec:ccd}. Left-multiplying (\ref{equ:multiPProof}) by
$\boldsymbol{R}^\dagger$ and then taking the Hermitian adjoint of both sides of
the resulting equation yields,
\begin{equation}
    \frac{1}{N\sqrt{L}} \boldsymbol{U} \boldsymbol{P}^\dagger \boldsymbol{R} =
    \boldsymbol{V} \boldsymbol{\Sigma}.
    \label{equ:multiPCCD}
\end{equation}
Taking the $k$th column of both sides of (\ref{equ:multiPCCD}) yields the $k$th
multi-observable degenerate CCD mode 
\begin{equation}
    \boldsymbol{V}_{k} = \frac{1}{N\sigma_k\sqrt{L}}\boldsymbol{U}
    \boldsymbol{P}^\dagger \boldsymbol{R}_{k}.
    \label{equ:kthCCD}
\end{equation}

Comparing (\ref{equ:kthePOD}) and (\ref{equ:kthCCD}), one sees that the $k$th
extended POD and degenerate CCD modes share much similarity. In particular,
since both $\boldsymbol{V}_{s}$ and $\boldsymbol{R}$ are unitary matrices of
size $L \times L$, $\boldsymbol{V}_{s,k}$ and $\boldsymbol{R}_k$ are of similar
forms. This shows that both modes can be written as a projection of
$\boldsymbol{U}\boldsymbol{P}^\dagger$ onto a unitary matrix of the same size.
However, since $\boldsymbol{V}_s$ is obtained by decomposing
$\boldsymbol{P}^\dagger$ while $\boldsymbol{R}$ by decomposing $\boldsymbol{P}
\boldsymbol{U}^\dagger/N\sqrt{L}$, in general, they are not the same. This
represents the key difference between the two methods, i.e. one uses an
energy-like rank in the subdomain only, while the other uses a correlation rank
involving both the subdomain and full domain. It is also this difference that
ensures the resulting CCD modes are orthogonal, while it is not necessarily so
for the extended POD.

However, in the special case where only one observable exists in the subdomains
$\boldsymbol{s}$ and only one extended POD mode results, both $\boldsymbol{V}_e$
and $\boldsymbol{R}$ reduce to $1$. Clearly, in this case, the $k$th extended
POD and degenerate CCD modes are identical, subject to a normalisation constant.
This also suggests that a key difference between the two is that an extra
dimension of time shift is allowed in CCD. It is in fact this difference that
results in a more effective order-reduction, which will be discussed in the
following sections. }



\add{In summary, one can see that CCD is different from the extended POD in the
following ways. First, CCD uses a norm involving both the subdomain and full
domain, while the extended POD uses a norm defined in a subdomain space.}
Second, it is not the energy of the flow within a subdomain that is maximised,
but the cross-correlation between the flow and the observable, which is the key
difference from the extended POD. Last, the matrix $\boldsymbol{P}$ is formed by
consecutively shifting the temporal delay between the flow and the observable.
This is why although $\boldsymbol{A}^\dagger \boldsymbol{A}$ can be written as
$\boldsymbol{U}(\boldsymbol{P}^\dagger \boldsymbol{P}) \boldsymbol{U}^\dagger$,
CCD is not weighted POD as $\boldsymbol{P}^\dagger \boldsymbol{P}$ is a
non-diagonal matrix formed by time shifting the observable, instead of a
diagonal weight independent of the flow variables. Note that $\boldsymbol{P}$
does not have to be within the flow field; instead, it can represent a variable
outside the flow field, a Fourier component of the flow, \add{a particular event
in a complex flow,} or an observable obtained by integrating the entire flow
field. 

\add{Apart from these differences, the connections between POD, extended POD and
CCD can also be shown. For example, mathematically POD can be regarded as a
special case of CCD when the observable is just an impulse exhibiting no
spectral preferences.} Specifically, if $p_{ij} =\delta_{i(N+1-i)}$ where
$i=1,2,\ldots, N$ and $\delta_{ij}$ is the Kronecker delta function, we see that
matrix $\boldsymbol{A}$ is a reversed $\boldsymbol{U}^\dagger$ and CCD reduces
to POD. Physically, this implies that $p$ contains identical frequency
components, and therefore exhibits no preferences in the spectral space.
Therefore, $\boldsymbol{U}$ is decomposed into modes ranked purely by their
energy. Similarly, mathematically CCD may reduce to the extended POD if
$\boldsymbol{P}$ is a mathematically constructed simple diagonal weight matrix
independent of any flow variables. The exact diagonal elements of course depend
on the specific subdomain to be interrogated in the extended POD.

\section{Validation}
\label{sec:validation}
\subsection{One-dimensional deterministic flow fields}
\label{subsec:1Ddetermin}
To validate that CCD can effectively extract flow events that correlate with
an observable, even under very low SNR, we create an artificial one-dimensional
unsteady flow field 
\begin{IEEEeqnarray}{rl}
    u(x, t) = &2 \cos(t -x )
    + 1.5 \cos(2t) \cos(2x)
    + \cos(3t) \cos(3x) \IEEEnonumber\\ 
	      &\negmedspace{} + 0.5\cos(4t)\cos(4x)
	      + \cos(6t)\cos(6x) \exp(-0.1(x-\pi)^2)
	      + 100r(t, x),
	      \label{equ:artificailFlow} \IEEEeqnarraynumspace
\end{IEEEeqnarray}
where $r(t, x)$ represents a random noise field  with a uniform probability
distribution over $[-0.5, 0.5]$, while other terms represent given flow
structures with different amplitudes. Note that the energy of the random noise
field is deliberately chosen to be {around $10^4$ times stronger} than the
defined flow structures. 

Suppose that $p$ represents an observable of interest at a specific point of the
flow field, for example, it may represent the skin friction fluctuations at one
point on the bottom wall within a turbulent channel flow. It is known that some
flow structures are the primary cause of the skin friction fluctuations while
others have minimal effects on them. Therefore, as an illustration we suppose
that $p$ is generated by the flow events represented by the first, second,
fourth and fifth terms \add{in (\ref{equ:artificailFlow})}, but not by the
third and last terms. For instance, $p$ may be given by 
\begin{equation}
    p(t) = \cos(t-\frac{\pi}{4}) + \sin(2t-\frac{\pi}{3}) + \cos(4t) +
    \cos(6t-\frac{\pi}{12}).
    \label{equ:artificialP}
\end{equation}
\add{Note that the amplitudes of the terms shown in (\ref{equ:artificialP}) are
chosen to be identical, although this is not at all necessary. In fact, they may
be changed arbitrarily without affecting the validity of the decomposition, for
instance, the amplitudes shown in (\ref{equ:artificailFlow}) can be used should
one be interested.}

\begin{figure}
    \centering
    \includegraphics[width=0.87\textwidth]{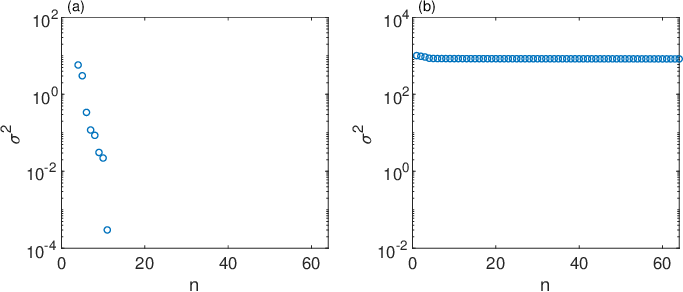}
    \caption{Comparison of the spectra of CCD (a) and POD (b) when $N=10^4$,
    $Q=128$ and $f_s^p = f_s^u=128/2\pi$. CCD is capable of effectively
    extracting the observable-correlated events leading to a low-rank spectrum,
    while POD results in a flat spectrum.}
    \label{fig:toy}
\end{figure}
Suppose that the flow field $u$ is sampled over $t\in [0, 2N\pi]$ at a sample
frequency $f_s^u = 128/2\pi$, where $N$ is an integer representing the number of
periodic cycles. Given the strong random noise in (\ref{equ:artificailFlow}),
$N$ is chosen to be a large number (only necessary when strong noise is
present). $p$, on the other hand, is sampled at the same sample frequency $f_s^p
= f_s^u$ but for a slightly longer duration of $2(N+1)\pi$. According to
section~\ref{subsec:ccd}, by choosing $\tau_1=0$ and $Q=128$, we can construct a
matrix $\boldsymbol{P}$ with $128$ rows straightforwardly. Within each snapshot,
the flow field is discretized on a mesh of $128$ points uniformly distributed
between [$0, 2\pi]$. In this example, $\Delta \tau = 2\pi/128$, therefore the
maximum frequency that can be resolved is limited by around $64/2\pi$.
Similarly, $Q=128$ implying that the frequency resolution is around $1/2\pi$. 

Following the procedures introduced in section~\ref{subsec:ccd}, both matrix
$\boldsymbol{U}$ and $\boldsymbol{P}$ can be easily constructed, where
$\boldsymbol{U}$ is of a size of $128 \times 128N$ while $\boldsymbol{P}$ is of
a size $128\times 128N$. Upon constructing the matrix $\boldsymbol{A}$, the CCD
can be carried out in a straightforward manner. \add{The resulting CCD spectrum,
i.e. the magnitude of the singular values against the mode number}, is shown in
figure~\ref{fig:toy}(a). To facilitate a direct comparison, the POD spectrum is
also shown in figure~\ref{fig:toy}(b). 
\begin{figure}
    \centering
    \includegraphics[width=0.87\textwidth]{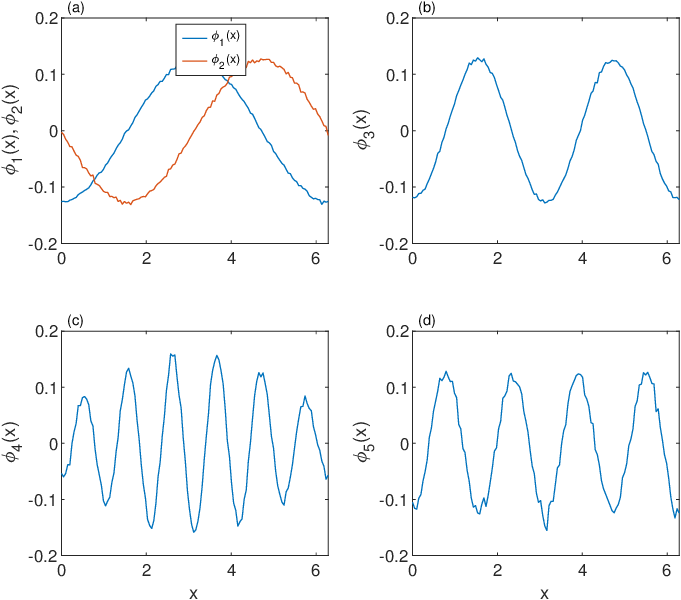}
    \caption{Extracted CCD modes when $N=10^4$, $Q=128$ and $f_s^p =
	f_s^u=128/2\pi$. They are most correlated with the observable $p$
	(corresponding to the \add{first, second}, fifth and fourth terms in
    (\ref{equ:artificialP}), respectively).} 
    \label{fig:modes}
\end{figure}

Figure~\ref{fig:toy} shows the CCD spectrum with a desired low-rank behaviour.
From figure~\ref{fig:toy}(a) we see that the five modes that correlate with the
observable can be robustly identified, even though the energy of the random
noise is \add{up to $10^4$ times stronger}.  Specifically, the first two
identical singular values form a pair, revealing a flow event of travelling-wave
nature, i.e. $\cos(t-x)$. The first mode of the pair corresponds to
$\sin(t+\phi)\sin(x+\phi)$ while the other to $\cos(t+\phi)\cos(x+\phi)$ ($\phi$
is an arbitrary phase delay), as demonstrated in figure~\ref{fig:modes}(a). The
third, fourth, and fifth singular values correspond to the flow events described
by the second, fifth, and fourth terms in (\ref{equ:artificailFlow}),
respectively. These can be confirmed by examining the corresponding mode vectors
shown in figure~\ref{fig:modes}(b-d). Most importantly, the $\cos(3x)$ mode,
which does not correlate with the observable, is robustly removed in the CCD
spectrum. This shows that CCD can effectively remove those uncorrelated flow
events while only keeping those correlated, and therefore works well for an
observable-targeted feature extraction and order reduction. 

The sixth to the ninth singular values ($\sigma_j^2$) shown in
figure~\ref{fig:toy}(a), which are two orders of magnitude weaker than the first
few modes, are artefacts introduced by the strong random noise. Note, however,
that these unphysical modes can be further suppressed robustly if the flow field
is sampled for a longer duration (larger $N$). All other values of $\sigma_j^2$
are below $10^{-24}$ and therefore not shown within the given range. As
discussed in section~\ref{subsec:physicialSignificance}, the singular values
represent the correlation strengths between corresponding CCD modes and the
observable. In this illustrative case, the observable is comprised of four modes
of equal amplitude, as shown in (\ref{equ:artificialP}), therefore the singular
values in figure~\ref{fig:toy}(a) are precisely the observable-correlated
fluctuation energy (subject to a fixed constant), as evidenced in
figure~\ref{fig:modes} (for example
$\sigma_1^2:\sigma_2^2:\sigma_3^2=2^2:2^2:1.5^2$).

On the other hand, figure~\ref{fig:toy}(b) shows that due to the strong random
noise the POD spectrum is completely corrupted and shown as a flat line. The
low-rank behaviour embedded within the data therefore cannot be identified. This
is expected, because POD modes are ranked by their corresponding fluctuation
energy. The random noise present in the flow field is \add{up to $10^4$
times stronger} than the observable-correlated events, and therefore completely
dominates the POD spectrum. More importantly, even though POD may be able to
extract the coherent structures when weaker noise is present, it cannot separate
the observable-correlated flow structures from those uncorrelated in the same
way as CCD does, since no information of $p$ is used. For example, the second
term of (\ref{equ:artificailFlow}) would stay in the POD spectrum and also
exhibit as a dominant mode.

\add{Having validated the decomposition, one can straightforwardly demonstrate
the effects of varying the sampling frequency, duration, time shift and
including multiple observables. The results agree well with the arguments
discussed in sections~\ref{subsec:resolution} and \ref{subsec:multiP}. For
conciseness, however, we do not include them in this section, but rather have it
shown in Appendix A.}

\subsection{One-dimensional statistical flow fields}
The example shown in figures~\ref{fig:toy} and \ref{fig:modes} illustrates the
capability of CCD in extracting flow events from highly noisy data. The
temporal signals given in (\ref{equ:artificailFlow}) are deterministic;
we can show in a similar manner that CCD can also effectively extract the
observable-correlated flow events when the temporal variation is statistical,
such as those exhibited in many turbulent flows. To show this, we construct an
artificial one-dimensional flow field
\begin{equation}
    u(x, t) = 3 s_1(t) \cos x
    + 2 s_2(t) \cos 3x
    + s_3(t) \cos 6x \exp(-0.1 (x-\pi)^2) + 10r(t, x),
    \label{equ:artificialFlowStatistical}
\end{equation}
where $s_i(t)$ ($i=1,2, 3$) represent three statistical processes. The $s_i(t)$
series are generated by a random number generator with different seeds in
MATLAB and then filtered using three different 6th-order Butterworth filters.
More specifically, $s_1(t)$ is filtered using a bandpass filter with lower and
upper cut-off frequencies of $0.2f_s$ and $0.4f_s$, respectively. The $s_2(t)$
and $s_3(t)$ series are filtered using low-pass filters with cut-off
frequencies of $0.2f_s$ and $0.15f_s$, respectively. For illustrative purposes,
we also add a random noise field that is two orders of magnitude more
energetic than $s_3(t)$. Suppose that the observable $p$ is generated by the
flow events represented by the second and third terms in
(\ref{equ:artificialFlowStatistical}), but not by the first, i.e.
\begin{equation}
    p(t) = s_2(t-\frac{\pi}{3}) + s_3(t) +2 \left[s_3(t)^2 - \overline{s_3(t)^2}\right] +
    3\left[s_3(t)^3 - \overline{s_3(t)^3}\right] + r(t).
    \label{equ:statisticalP}
\end{equation}
Note that because the observable may be non-linearly related to the flow
dynamics, we also add in (\ref{equ:statisticalP}) two nonlinear terms of
$s_3(t)$, as shown by the two bracket terms. Similarly, the observable may be
also subject to noise contamination. A statistical random noise $r(t)$, with a
uniform distribution over $[-0.5, 0.5]$, is therefore also added. The flow field
is again sampled at $f_s^u=128/2\pi$ on a uniform spatial mesh of 128 points
over the time interval $[0, 2N\pi]$, while $p$ is sampled over $[0, 2(N+1)\pi]$
using the same frequency $f_s^p=128/2\pi$. 

\begin{figure}
    \centering
    \includegraphics[width=0.87\textwidth]{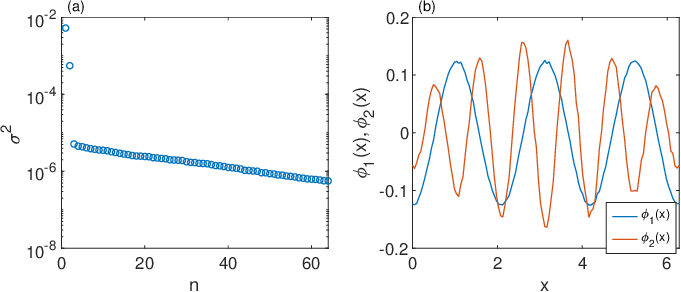}
    \caption{The CCD spectrum (a) and the first and second modes (b). $N$ takes
    the value of $10^4$ \add{but $10^3$} may also be used at the expense of
    convergence.}
    \label{fig:statisticalToy}
\end{figure}
Routine use of the decomposition yields the CCD spectrum and the first two
modes, as shown in figures~\ref{fig:statisticalToy}(a) and
\ref{fig:statisticalToy}(b), respectively. Clearly, the leading-order mode
corresponds to the second term in (\ref{equ:artificialFlowStatistical}), while
the second-order mode the third. This can be clearly seen from
figure~\ref{fig:statisticalToy}(b). It is worth noting that the observable also
contains the square and cube of $s_3(t)$, but this does not appear to affect the
identification of the second mode. Indeed, CCD works by maximizing the
correlation between the flow field and the observables, but in general it does
not limit the observable being a linear function of the flow field.
Additionally, the first term of (\ref{equ:artificialFlowStatistical}), due to it
being uncorrelated with $p$, is effectively removed in the CCD spectrum. Other
higher-order modes are more than two orders of magnitude lower than the first
two. Again, as $N$ increases, these unphysical modes can be further suppressed,
while the physical modes resolved more accurately. Note that in this
illustrative example, the observable $p$ is also corrupted by the random noise,
but CCD continues to work robustly.

\section{Applications to numerical and experimental data}
\label{sec:realFlow}
\add{Having validated the method, in this section CCD is used to decompose
numerical and experimental data in order to demonstrate its potential use in
fluid mechanics. Three flows are used, including a turbulent channel flow, a
subsonic jet and a wake flow past a cylinder. Where possible, POD results are
also included for comparison. In all cases, the simple $L_2$ norm of the flow
$\boldsymbol{u}_i$ is used in POD.}

\subsection{Turbulent channel flow}
As an illustrative example, we first apply CCD to a Direct Numerical Simulation
(DNS) database of turbulent channel flows.  \add{The database was obtained from
a turbulent channel flow using the code developed by \citet{Lee2015}.} The
computational domain is of $4\pi H \times 2H \times 2\pi H$ in the streamwise
($x$), wall-normal ($y$) and spanwise ($z$) directions, respectively, where $H$
denotes the half-height of the channel. \add{The domain is discretized using
$192$, $128$ and $192$ points in $x$, $y$ and $z$ directions, respectively.} The
friction Reynolds number $Re_\tau$ defined as $\rho u_\tau H/\mu$, where $\rho$,
$\mu$ and $u_\tau$ denote the fluid density, dynamic viscosity and friction
velocity at the wall respectively, is around $180$. \add{The time step is fixed
at $0.01H/U_b$, where $U_b$ is the bulk flow velocity. The flow is sampled every
$100$ time steps, resulting in a sampling frequency of $f_s^u = U_b/H$.} In
total, $1687$ snapshots of the flow field are recorded.

In turbulent channel flows, skin friction represents a significant operational
cost in applications such as long-range oil transport~\citep{kim2011physics}.
The control of turbulent skin friction is therefore of particular interest and
has been studied extensively in the literature~\citep{gad2007flow}. To
understand the physical mechanism concerning its generation and suppression, it
is crucial to extract the turbulent flow structures that determine the skin
friction. CCD is therefore suitable for such a diagnosis. As mentioned in
section~\ref{sec:ccd}, without the data storage limit, the observable is allowed
to be sampled at a much higher frequency than the flow field. In this example,
the sampling frequency $f_s^p = 10 f_s^u$, resulting in an interval of
$\Delta\tau = 0.1H/U_b$ and $16870$ samples for the skin friction.
\begin{figure}
    \centering
    \includegraphics{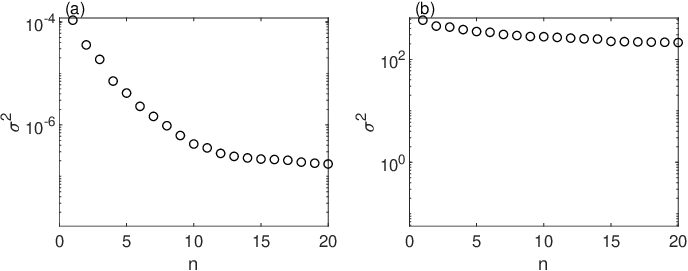}
    \caption{The spectra of (a) CCD and POD; The CCD spectrum exhibits a much
    steeper decay as mode number $n$ increases, indicating a more effective
    order reduction.}
    \label{fig:Spectra}
\end{figure}
\begin{figure}
    \centering
    \includegraphics[width=0.89\textwidth]{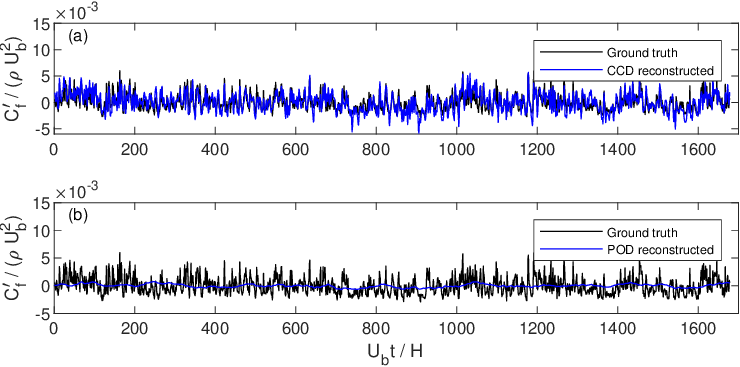}
    \caption{Reconstruction of the dimensionless wall friction coefficient using
    the first 6 CCD (a) and POD (b) modes, respectively.}
    \label{fig:reconstructionChannel}
\end{figure}
\begin{figure}
    \centering
    \includegraphics[width=0.95\textwidth]{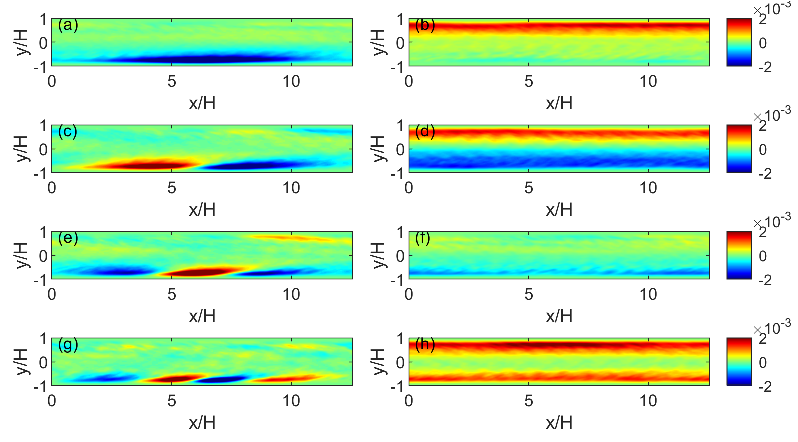} 
    \caption{Front views of the first four CCD (a,c,e,g) and POD (b, d, f, h)
    modes. The $z$ coordinate is fixed at $\pi H$ for (a, c, e, g). The spanwise
    widths of the streaks in (a,c,e,g) are around $0.2H$.}
    \label{fig:pcdMode}
\end{figure}

\add{Considering the statistical homogeneity of the flow, we use the skin
friction sampled at $x=2\pi H$ and $z=\pi H$ on the lower wall ($y=-H$) as the
observable and choose the streamwise velocity as the flow variable in order to
extract the coherent structures.} Considering the short temporal correlation
scale, we choose $Q=100$ and $\tau_1 = -50\Delta \tau$. Using the procedures
described in section~\ref{sec:ccd}, we perform CCD and obtain the resulting
singular values and CCD modes. The singular values are shown in
figure~\ref{fig:Spectra}(a). \add{Also shown is the spectrum from POD in
figure~\ref{fig:Spectra}(b), where the streamwise velocity is decomposed}.
Comparing the two we see that the CCD spectrum is markedly different from that
of POD. In particular, the CCD spectrum exhibits a much quicker decay. For
example, higher-order modes ($\ge5$) are one order of magnitude lower, whereas
the POD spectrum is rather flat. This signals a quicker reconstruction of the
skin friction using CCD modes. Indeed, using the first $6$ modes recovers more
than $80\%$ of the total skin friction at the observer point, as shown in
figure~\ref{fig:reconstructionChannel}(a). \add{The high-frequency deviation may
be further reduced if the observable is allowed to be sampled faster.} In
contrast, the first $6$ POD modes only recover less than $5\%$ energy, as shown
in figure~\ref{fig:reconstructionChannel}(b).

The resulting CCD modes are shown in figure~\ref{fig:pcdMode}. We see that the
CCD modes take the form of streamwise streaks slightly above the bottom wall, in
accordance with current understanding. More importantly,
figure~\ref{fig:pcdMode} also shows that they are spatially localized around the
observer point. This is particularly true in the spanwise direction with a
streak width of less than $0.2H$. Moreover, higher-order modes have increasingly
short spatial and temporal scales. To the best knowledge of the authors, such a
quantitative and unambiguous characterization of these structures specifically
targeting the skin fluctuation in the middle of the wall has not been reported
in the literature. In contrast, although the POD modes take the form of streaks,
they are not localized around the observer point, but stretched in the
streamwise direction and scattered in the spanwise direction instead.  Moreover,
the first few modes do not exhibit a clear decrease of either spatial or
temporal scales, signalling a slower reconstruction of the skin friction.

\add{Although CCD focuses on examining the flow structures that contribute to
the skin friction at one individual point, it in fact does not lose generality.
This is because the flow is homogeneous in the streamwise direction; the
structures that generate the skin friction at other locations on the wall remain
identical (subject to a shift in space). However, by focusing on the observable
at a specific point, one expects to obtain a more effective order reduction
since our interest is more focused. The fact that the flow is homogeneous can
also be exploited to improve the convergence of the resulting flow. Instead of
using the observable at one point, one can use multiple points along different
spanwise or streamwise locations. They can be treated as independent
realisations, with which the resulting mode indeed converges better. However,
since the structures remain similar to those in figure~\ref{fig:pcdMode}, we
omit showing their contours repetitively.} 

\begin{figure}
    \centering
    \includegraphics[width=0.9\textwidth]{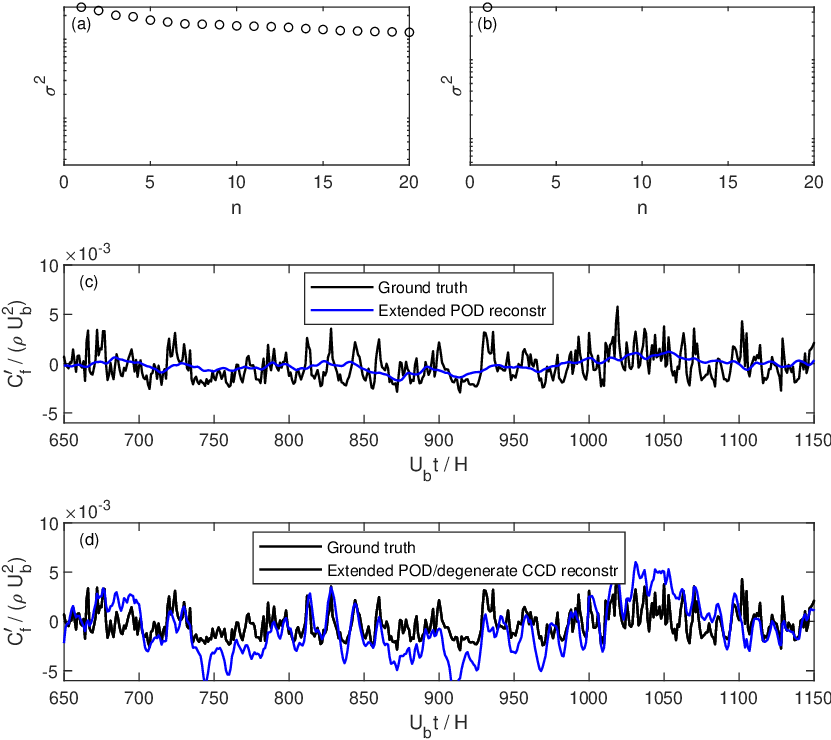}
    \caption{\add{The spectrum of extended POD using (a) a subdomain of size
    $4\pi H \times 2\pi H$ and (b) a subdomain consisting of the observable
    point only; Reconstructed skin friction fluctuations using (c) a subdomain
    of size $4\pi H \times 2\pi H$ (using the first $6$ extended POD modes ) and
    (d) a subdomain of only one point (only $1$ mode results).}}
    \label{fig:extendPOD}
\end{figure}
\add{Note that part of the reason why the skin friction reconstruction using POD
is slow is due to its use of energy within the entire domain as the norm. Since
the extended POD can be used to target more towards the observable, it is
interesting to compare it with CCD in detail. To show this, we first perform the
extended POD using skin friction on the wall.}
\add{The resulting singular values and reconstruction of the skin friction using
the first 6 extended POD modes are shown in figures~\ref{fig:extendPOD}(a) and
\ref{fig:extendPOD}(c), respectively. The resulting spectrum of singular values
exhibits a similar slow decay to that shown in POD. This is consistent with a
similar reconstruction of the skin friction, as shown in
figure~\ref{fig:extendPOD}(c), where a limited time range from $650$ to $1150$
is shown for clarity. However, comparing to
figure~\ref{fig:reconstructionChannel}(b), the skin friction reconstruction
appears slightly improved when the wall shear stress is used as the subdomain in
the extended POD.}


\add{One is, therefore, interested in seeing how much the reconstruction can
improve by using increasingly small subdomains centring around the observable.
In the ultimate case, the subdomain can be chosen to consist of the observable
point only. We choose to perform extended POD using such a special subdomain.
Note that this is identical to the degenerate CCD where no time shift is
included between the observable and the flow. We expect the resulting mode to
better target the observable, which is indeed the case, as shown in
figure~\ref{fig:extendPOD}(d). The extended POD modes can capture an overall
trend in the skin friction variation. However, it is important to note that
since there is only one mode available, as can be seen in its spectrum shown in
figure~\ref{fig:extendPOD}(b), this is the best reconstruction one can achieve
using the extended POD. }

\add{On the other hand, since the decomposition is also a degenerate case of
CCD, this represents the worst reconstruction one would obtain using CCD.
Indeed, by including the dimension of time shifts, $\boldsymbol{P}$ would have a
rank of more than 1, and the reconstruction using CCD improves considerably, as
shown in figure~\ref{fig:reconstructionChannel}(a). Note that the reconstruction
further improves as one includes more CCD modes, the family of which forms a
complete orthonormal set. In addition, figure~\ref{fig:extendPOD}(d) shows that
only an overall trend of the skin friction is captured in the reconstruction,
and the deviation occurs mainly in the high-frequency regime. This is expected,
since this degenerate CCD corresponds to a sampling interval $\Delta\tau=\infty$
for the observable, hence a failure to resolve high-frequency components.}

\add{In summary, using only one point where the observable is located in the
extended POD better targets the observable, but the resulting one mode limits
the capability of separating multiple flow structures that possibly coexist
within the flow. To do that, a sufficiently large region is preferred,
compromising the observable specificity. This appears a trade-off between
targeting a local observable and separating multiple flow structures. CCD does
not have this limitation, and this relaxation is enabled by exploiting the
“hidden” time-shift dimension. This reflects a key difference between CCD and
the extended POD. More importantly, this also adds the flexibility of fully
exploiting a different (possibly much higher) sampling frequency.}

Figures~\ref{fig:Spectra} and \ref{fig:pcdMode} show that CCD works well in
extracting the coherent structures that are most correlated with the given
observable. This is further evidenced by a quick reconstruction of the skin
friction using the first few CCD modes. Note again that in this example the
observable is sampled at a much higher frequency than the flow. This
flexibility, as mentioned in section~\ref{subsec:connectionPOD}, \add{plays an
important role in the successful feature extraction and order reduction.}

\subsection{Turbulent subsonic round jets}
In this section, we apply CCD to a numerical dataset of a turbulent subsonic
round jet. Using the pressure fluctuations as the observable we can examine the
flow structures that are most correlated with them. Directly resolving far-field
pressure fluctuation is rarely possible in numerical turbulence simulations,
hence in this example we examine the near-field pressure fluctuation instead.
\add{This is expected to suffice for the purposes of demonstrating the potential
use of CCD.} \add{The near-field dynamics of turbulent jets is expected to
connect with their mixing and acoustic characteristics, and is therefore studied
extensively in the literature. Order reduction techniques are widely used. This
includes POD or the extended POD with a variety of
norms~\citep{Freund2009,Sinha2014,Schmidt2019}, the resolvent/input-output
analysis~\citep{Jeun2016,Pickering2021,Bugeat2024} and other source
identification methods that we do not aim to show exhaustively. Moreover, the
near-field pressure fluctuations are crucial in determining installed jet
noise~\citep{Lyu2017a,Lyu2019a}, therefore its modelling and control have
practical uses.}




The numerical data is extracted from an earlier work~\citep{Lyu2017a}, where an
LES simulation of a subsonic round jet was performed. Only a slice of data on
one azimuthal plane is used, but it should be sufficient for illustration
purposes. The jet Mach number is $M_j = 0.5$ while the nozzle diameter $D$ is
$2$ inches.
The computational domain is axisymmetric, with the streamwise coordinate $x$
extending from $0$ to $20D$ and lateral coordinate $r$ extending to $4D$. The
computational domain is discretized using $512$ and $97$ points in the $x$ and
$r$ directions, respectively.

We choose the near-field pressure fluctuation at $x/D=10$ and $r/D=4$ as the
observable. At this close distance, the observable is likely to include both
acoustic and hydrodynamic pressure fluctuations. \add{In addition, we choose the
pressure field as the flow variable to be decomposed. The same is used in a
reference POD decomposition.} The flow is sampled at a frequency of $f_s^u =
4U_j/D$ for a duration of $200 D/U_j$, where $U_j$ is the jet exit velocity. The
near-field pressure $p$ is sampled at the same frequency but for a longer
duration of $280 D/U_j$. This results in a $Q$ value of $320$. Due to the short
distance between the flow field and the near-field pressure fluctuations, we
choose the time delay $\tau_1$ to be 0.
\begin{figure}
    \hspace{0.5cm}
    \includegraphics[width=0.85\textwidth]{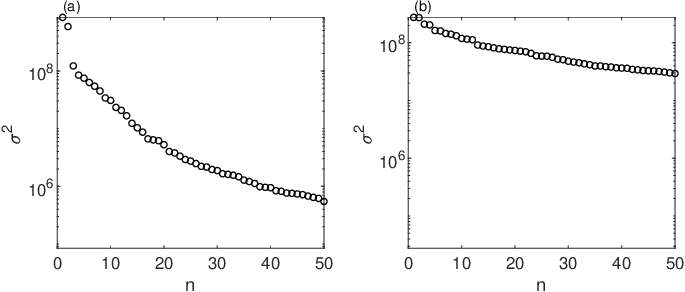}
    \caption{The CCD (a) and POD (b) spectra of the unsteady pressure field on
	a $x-r$ plane. The CCD spectrum shows a clear low-rank behaviour compared
    to POD.}
    \label{fig:jetpcdspectrum}
\end{figure}
\begin{figure}
    \centering
    \includegraphics[width=0.96\textwidth]{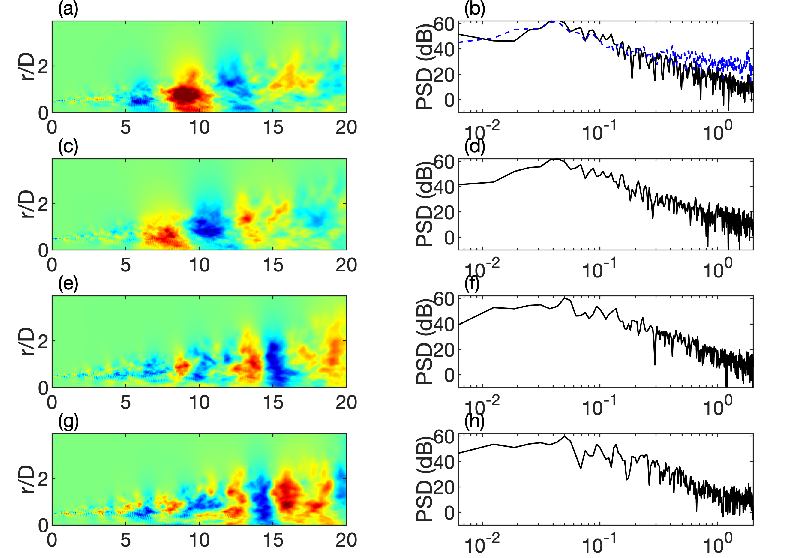}
    \caption{The first 4 CCD modes (a,c,e,g) and \add{PSD spectra (b,d,f,h) of
    their corresponding temporal coefficients, where the blue dashed line in (b)
    represents the spectrum of the observable with its magnitude scaled for an
    easier comparison;} mode 1 (a-b), mode 2 (c-d), mode 3 (e-f), mode 4 (g-h).
    As the mode number increases the CCD modes are characterised by increasingly
    short spatial scale and high frequency components.}
    \label{fig:jetCCDModes1}
\end{figure}
With the procedure described in section~\ref{sec:ccd}, the CCD spectrum is
shown in figure~\ref{fig:jetpcdspectrum}. Also shown is the POD spectrum to
facilitate a direct comparison. Only the first $50$ singular values are shown.
Compared to POD, the CCD spectrum exhibits a more rapid decay as the mode
number $n$ increases. In particular, at small mode numbers the CCD spectrum
shows a clear low-rank behaviour. The first two modes are almost one order of
magnitude stronger than higher-order modes. This is in direct contrast to the
POD spectrum, where the low-rank behaviour is not pronounced.

\begin{figure}
    \centering
    \includegraphics[width=0.96\textwidth]{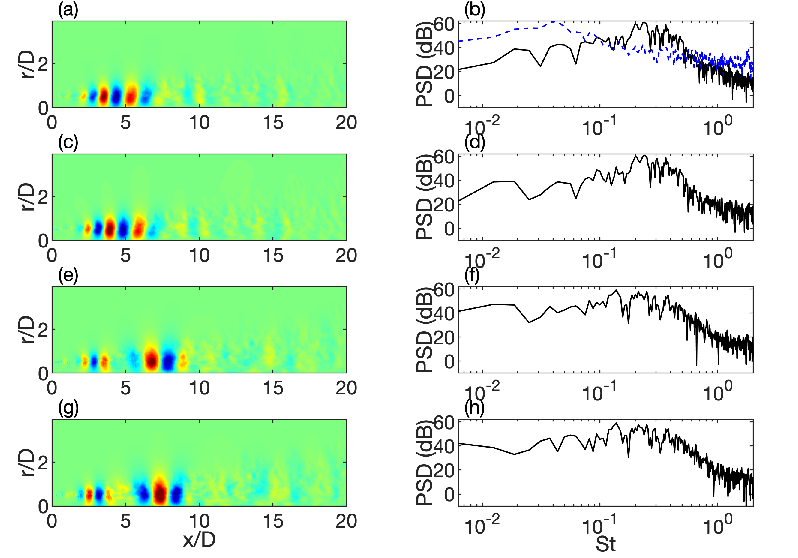}
    \caption{The first 4 POD modes (a,c,e,g) and \add{PSD spectra (b,d,f,h) of
    their corresponding temporal coefficients, where the blue dashed line in (b)
    represents the spectrum of the observable with its magnitude scaled for an
    easier comparison;} mode 1 (a-b), mode 2 (c-d), mode 3 (e-f), mode 4 (g-h).
    As the mode number increases the POD modes have larger spatial scales with
    more low-frequency components.}
    \label{fig:jetPODModes1}
\end{figure}

This can be understood from figures~\ref{fig:jetCCDModes1} and
\ref{fig:jetPODModes1}, where the first 4 CCD and POD modes
$\phi_k(\boldsymbol{x})$ and the corresponding Power Spectral Densities (PSDs)
of their temporal expansion coefficients $a_k(t)$ are shown, respectively.
Clearly, the first two CCD modes are large flow structures exhibiting relatively
low-frequency behaviour, whereas the leading-order POD modes have much shorter
scales with a well-known dominant frequency at around $St=0.3$, where $St$ is
the Strouhal number defined using $U_j$ and $D$. \add{The PSD spectrum of the
observable is also included with its magnitude scaled for an easier comparison.}
Since the observer is located at $x/D=10$ and $r / D =4$, the pressure
fluctuations inevitably include the signatures of the downstream large coherent
structures. The similar first two singular values shown in
figure~\ref{fig:jetpcdspectrum} and similar mode shapes shown in
figure~\ref{fig:jetCCDModes1} indicate a convection behaviour of this large
structure. CCD decomposition can take this into consideration and yield an
observable-relevant low-frequency fluctuation mode. The leading-order POD modes,
on the other hand, are ranked only by the fluctuation energy and, therefore, are
not as relevant as the CCD modes.  

\begin{figure}
    \centering
    \includegraphics[width=0.87\textwidth]{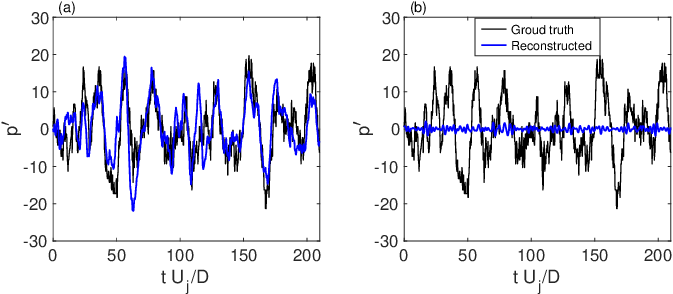}
    \caption{The reconstructed pressure fluctuations at the observable location
    using the first $2$ CCD (a) and POD modes (b), respectively.}
    \label{fig:reconstruction}
\end{figure}

Note that the singular value represents a measure of the cross-correlation in
the $L_2$ norm and, therefore, is in general not equal to the energy of the CCD
modes contained in the flow, nor is it equal to the energy of the corresponding
correlated component in the observable. Nevertheless, since the decomposition
targets more at the observable, we expect that it can reconstruct the pressure
fluctuations at the observable position using much fewer modes. \add{This is
indeed the case, as shown in figure~\ref{fig:reconstruction}, where only the
first two CCD and POD modes are included to calculate the reconstructed pressure
fluctuations at the observable position, respectively. As can be seen from
figure~\ref{fig:reconstruction}(a), using two CCD modes can yield a good
reconstruction, which is in contrast to POD shown in
figure~\ref{fig:reconstruction}(b).}
Note that in this application we use the pressure field as the flow $u$, whereas
in general a combined velocity and pressure field may be used. We can show that
a similar result may also be obtained when a combination of velocity and
pressure fluctuations is used in CCD. 

At large mode numbers, the CCD spectrum shows a steeper decay, and higher-order
modes tend to have increasingly short scales together with higher frequencies,
as shown in figures~\ref{fig:jetpcdspectrum} and \ref{fig:jetCCDModes1},
respectively. Note that the singular values represent the correlation strength
between the CCD modes and the observable, therefore the decay of the singular
values is determined by both the energy of the flow and the observable and the
coherence decay between them. Therefore, the steeper CCD spectrum suggests that
although the pressure fluctuations consist of energetic structures of various
scales, they may not be equivalently important in contributing to the
observable, therefore the coherence between the two may decrease rapidly. On the
other hand, the POD spectrum decays much more slowly, and as the mode number
increases the POD mode starts to capture more downstream large structures with
more low-frequency content, as shown in figure~\ref{fig:jetPODModes1}. That the
POD spectrum decays more slowly is attributed to the fact that the POD spectrum
is determined solely by the energy of flow and, therefore, does not depend on
its coherence with the observable.



Figure~\ref{fig:jetCCDModes1} shows that each CCD mode corresponds to a unique
temporal variation. Unlike the Fourier analysis, each of these temporal
variations is spectrally broadband. In essence, CCD decomposition works as a
special spectral transform of the flow based on its correlation with the
observable. Note, however, that the sampling frequency and duration are limited
in this simulation, and further analysis using longer samples is needed for
better statistical convergence. In addition, due to current data availability,
we only decompose the \add{near-field} pressure, and it would be interesting to
apply this technique to extract acoustically dominant flow features in future
studies. Nevertheless, it suffices for the purpose of demonstrating the
potential application of CCD. 

\subsection{Unsteady wake flows over cylinders}
\begin{figure}
    \centering
    \includegraphics[width=0.9\textwidth]{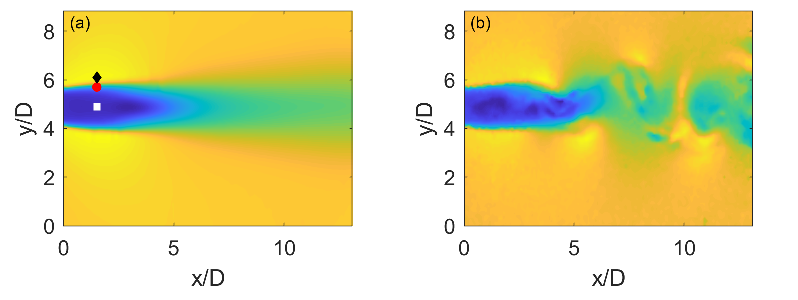}
    \caption{The mean (a) and instantaneous (b) streamwise velocity
    distributions of the immediately downstream wake over a cylinder
    flow~\citep{Renn2023}. The black diamond, red circle and white square are
    located at $x/D=1.5$ but $y/D=6.1$, $y/D=5.7$, and $y/D=4.9$, respectively.}
    \label{fig:pivsnapshot}
\end{figure}

In this example, we apply CCD to the experimental data of an unsteady wake flow
behind a cylinder. The experiment was performed in a water tunnel using the
two-dimensional time-resolved Particle Image Velocimetry (PIV) technique. The
cylinder had a diameter of $D=9.53$ mm while the Reynolds number was fixed at
$650$. The interrogation window was a rectangle immediately behind a cylinder in
the wake and measured $13D \times 9D$ in the streamwise ($x$) and cross-stream
($y$) directions, respectively. Details of the experimental setup can be found
in \citet{Renn2023}. The velocity field was sampled at a frequency of around
$50$ Hz on a mesh of $N_x=133$ and $N_y=89$, and in total $N=8250$ snapshots
were obtained. The mean and instantaneous streamwise velocity fields are shown
in figure~\ref{fig:pivsnapshot} for reference. As shown in
figure~\ref{fig:pivsnapshot}, the mean flow exhibits the expected symmetry
across the wake, while the instantaneous velocity field shows a clear vortex
shedding behaviour behind the cylinder.
\begin{figure}
    \hspace{0.5cm}
    \includegraphics[width=0.85\textwidth]{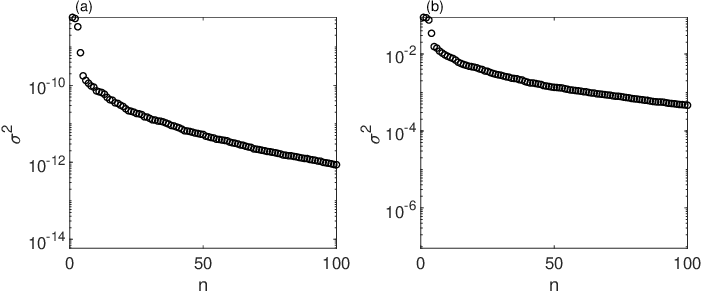}
    \caption{The CCD (a) and POD (b) spectra of the streamwise velocity
	fluctuations on an $x-y$ plane with the cross-stream velocity fluctuation at
    $y/D=5.7$ as the observable.} 
    \label{fig:pivpcdspectrum}
\end{figure}
The vortex shedding occurring when the Reynolds number exceeds a critical
number is one iconic feature of the flow over cylinders. Given its wide
applications such as wind blowing over chimneys and high-rise buildings, its
control has attracted significant attention in the fluid mechanics
community~\citep{Choi2008}. Many techniques exist, including both passive and
active controls. Earlier studies show that using the feedback signal recorded
in the wake, vortex shedding can be successfully suppressed or even eliminated
at low Reynolds numbers~\citep{Williams1989, Roussopoulos1993, Park1994}. In
designing a closed-loop active control system such as the one in
\citet{Park1994}, one primary interest is to identify the optimal location to
place the feedback sensor. Ideally, the observable, such as the cross-stream
velocity, at the feedback sensor location should maintain a strong correlation
with the vortex structures shed from the cylinder. CCD may be used to give an
initial assessment of the correlation between the sensed signal and the vortex
structures.

\begin{figure}
    \centering
    \includegraphics[width=0.9\textwidth]{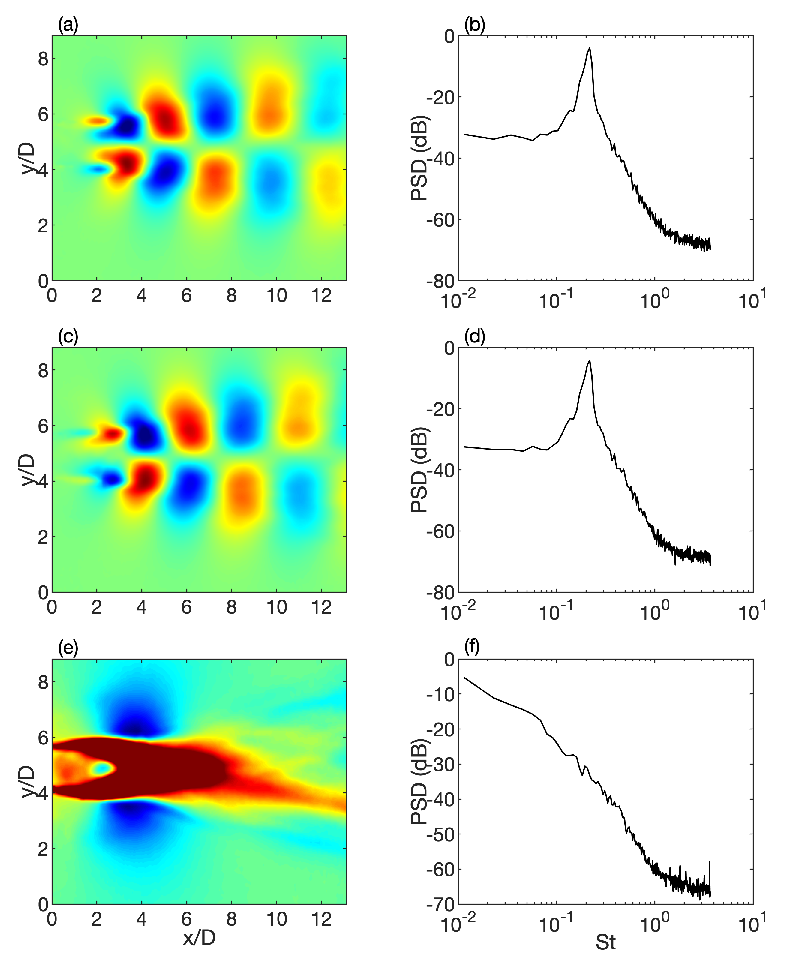}
    \caption{The first 3 CCD modes (a,c,e) and \add{PSD spectra (b,d,f) of their
    corresponding temporal coefficients;} mode 1 (a-b), mode 2 (c-d) and mode 3
    (e-f). }
    \label{fig:pivCCDModes1}
\end{figure}
\begin{figure}
    \centering
    \includegraphics[width=0.9\textwidth]{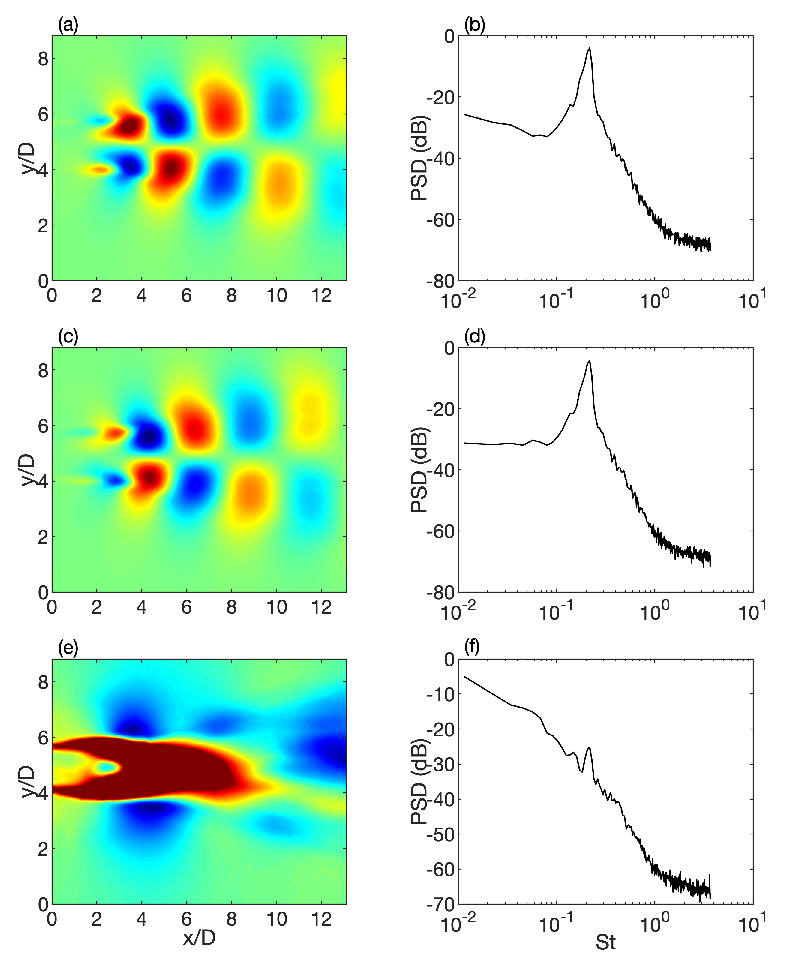}
    \caption{The first 3 POD modes (a,c,e) and \add{PSD spectra (b,d,f) of their
    corresponding temporal variation coefficients;} mode 1 (a-b), mode 2 (c-d)
    and mode 3 (e-f). }
    \label{fig:pivPODModes1}
\end{figure}
As an illustration, we choose the observable to be the cross-wake
velocity~\citep{Park1994} in the initial shear layer behind the cylinder, for
example at $x/D=1.5$ and $y/D=5.7$ as shown by the red circular dot in
figure~\ref{fig:pivsnapshot}(a). \add{Using this observable, we may decompose
the streamwise velocity field using CCD. Again, the streamwise velocity is
correspondingly used in POD.} The time shift $\tau_1$ is chosen to be
$-4Q\Delta\tau/5$ while $Q$ is chosen to be $N/3$.
Figures~\ref{fig:pivpcdspectrum}(a) and (b) show the CCD and POD spectra,
respectively. Clearly, both CCD and POD capture the dominant vortex shedding
behaviour, and the two nearly identical singular values reflect a convecting
behaviour of the shed vortices. Figures~\ref{fig:pivCCDModes1} and
\ref{fig:pivPODModes1} show the corresponding first three CCD and POD modes and
their corresponding PSDs, respectively. Clearly the first two vortex shedding
modes from both CCD and POD are virtually identical, which can be seen from both
the mode shape and their corresponding PSD spectra. The CCD spectrum shows a
slightly smaller singular value for the third mode, which is somewhat more
symmetric, whereas the similar mode resulting from POD obtains a similar
singular value compared to the leading-order mode. This suggests that although
this mode carries one of the largest energy, it is slightly less correlated with
the cross-stream velocity fluctuation at $x/D=1.5$ and $y/D=5.7$. 
\begin{figure}
    \hspace{0.5cm}
    \includegraphics[width=0.85\textwidth]{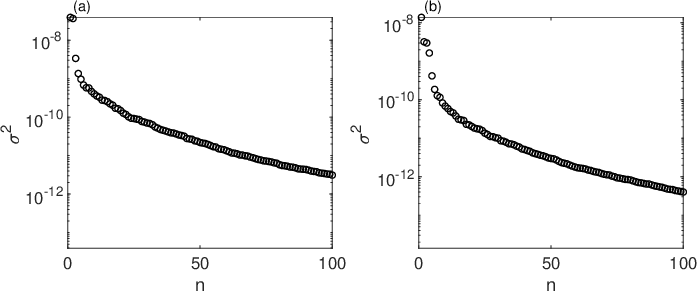}
    \caption{The CCD spectra of the streamwise velocity fluctuations when the
	observable is located at $y / D = 4.9$ (a) and $y / D=6.1$ (b),
    respectively.} 
    \label{fig:pivpcdspectrumaway}
\end{figure}

If we keep $x/D=1.5$ but move the observable position further away from the
shear layer, for example, to $y/D=4.9$ and $y/D=6.1$ as shown by the white
square and black diamond symbols respectively in figure~\ref{fig:pivsnapshot},
these three modes can still be identified using CCD, but their relative singular
values changed significantly, as shown in figure~\ref{fig:pivpcdspectrumaway}.
This implies that these modes correlate differently to different observables.
Evidently, the first and second modes in figure~\ref{fig:pivpcdspectrumaway}(a)
represent the vortex shedding modes. Their mode shapes are similar to those
shown in figure~\ref{fig:pivCCDModes1}(a) and (b), so we omit a repetitive
presentation.

However, it is important to note that these singular values are much larger
compared to those shown in figure~\ref{fig:pivpcdspectrum}(a), suggesting they
are more strongly correlated with the observable. More importantly, the third
singular value drops rapidly, almost one order of magnitude weaker than the
leading-order mode. From the feedback control point of view, this would be a
good candidate for placing the feedback sensor owing to its higher correlation
with our interested flow events and simultaneously a higher SNR.
Figure~\ref{fig:pivpcdspectrumaway}(b) shows that the singular values
corresponding to the vortex shedding modes are slightly lower than those shown
in figure~\ref{fig:pivpcdspectrum}(a) with an even stronger leading-order
non-shedding mode. Consequently, this would be a position to be avoided for
placing the feedback sensor. This may be why the wake centreline was used to
place the feedback sensors in \citet{Park1994}.

\section{Conclusion}
A data-driven method referred to as CCD is proposed in this paper to decompose
complex flows into modes ranked by their correlation strength with an
observable. The method is based on the canonical correlation analysis in
classical statistics. The method is validated for both deterministic and
statistical flow events. First, the results show that CCD can effectively
extract the observable-correlated flow features while suppressing those
uncorrelated in both cases. CCD, therefore, results in more low-rank spectra
compared to POD. Second, CCD can effectively extract those observable-correlated
flow structures even under low SNRs. Third, numerical validation shows that the
sampling frequency and duration of the observable determine the frequency limit
and resolution while that of the flow are to ensure the convergence of the
cross-correlation. Longer sampling of the flow and including multiple
observables can improve the convergence of the resulting CCD modes. Therefore,
CCD is particularly suitable for experimental data because long samples can be
more conveniently obtained. Lastly, as no linearity is assumed, CCD is capable
of extracting nonlinear flow events similar to POD, provided a non-negligible
correlation exists between the flow and the observable.  

As an illustrative example, the method is first used to analyse a turbulent
channel flow obtained using DNS. The flow structures that are most correlated
with the skin friction at the point in the middle of the bottom wall are
extracted. \add{It is shown that CCD yields a spectrum of singular values that
decays rapidly as the mode number increases compared to POD.} The first 6 CCD
modes effectively recover more than $80\%$ of the skin friction fluctuations.
\add{The extended POD using only one observable point can better target the
observable, and is found to be equivalent to the degenerate case of CCD when no
time shift between the flow and observable is used.} The CCD modes take the form
of streamwise streaks slightly above the wall. More importantly, the streamwise
and spanwise extent of these streaks are unambiguously determined. As the mode
number increases, CCD modes have increasingly short spatial and temporal scales. 

In a subsequent example, CCD is used to decompose the unsteady pressure field of
a turbulent subsonic jet using a near-field pressure fluctuation as the
observable. Results show that CCD results in a steeper spectrum compared to POD.
In particular, the CCD spectrum exhibits a clear low-rank behaviour and the
corresponding modes correspond to the large coherent flow structures that
convect downstream. The first two CCD modes recover $80\%$ of the energy of the
near-field pressure fluctuations. The method is subsequently applied to analyse
the unsteady vortex shedding behind a cylinder. It shows that similar modes to
POD can be robustly identified, but their strengths depend crucially on the
observable and its locations, suggesting that these modes correlate differently
with observables at different locations. This diagnosis would be useful for
determining the optimal location for placing the feedback sensor in a
closed-loop control of the vortex shedding behind a cylinder.

Note that the examples shown in the paper are only for illustrations. They
suffice for the purpose of demonstrating the potential uses of CCD, but further
improvements are needed for a more in-depth analysis. For example, we can see
that both the sampling frequency and sampling duration in the jet example are
rather limited; \add{therefore, a faster sampling of the observable, together
with a longer sampling duration, is needed for more accurate diagnosis. In
addition, a possible far-field noise diagnosis using pressure or the Lighthill
stress tensors as the flow variables may be conducted. These, together with the
application of CCD in the spectral space such as that shown in
section~\ref{subsec:ccd}, form some of our future work.} 

\vspace{-0.2cm}
\section*{Acknowledgements} 
The author wishes to thank Prof. Ann Dowling and
Prof. Tim Colonius for the stimulating discussions on including multiple
observables and the low-rank properties of CCD modes. The author would like to
gratefully thank Prof. Jie Yao for sharing the DNS data of channel flows. The
author is very grateful to Dr. Cong Wang and his collaborators for agreeing to
use their PIV data on cylinder wakes. The author wishes to thank Dr. I. Naqavi
who performed the LES simulations in our last collaborative
publication~\citep{Lyu2017a}, from which part of the data is extracted and
reused. The author also wishes to gratefully acknowledge the National Natural
Science Foundation of China (NSFC) under the grant number 12472263.

\textbf{Declaration of interests. The authors report no conflict of interest.}

\vspace{-0.35cm}
\appendix
\add{\section{The effects of sampling frequency, duration and multiple observables}

\begin{figure}
    \centering
    \includegraphics[width=0.87\textwidth]{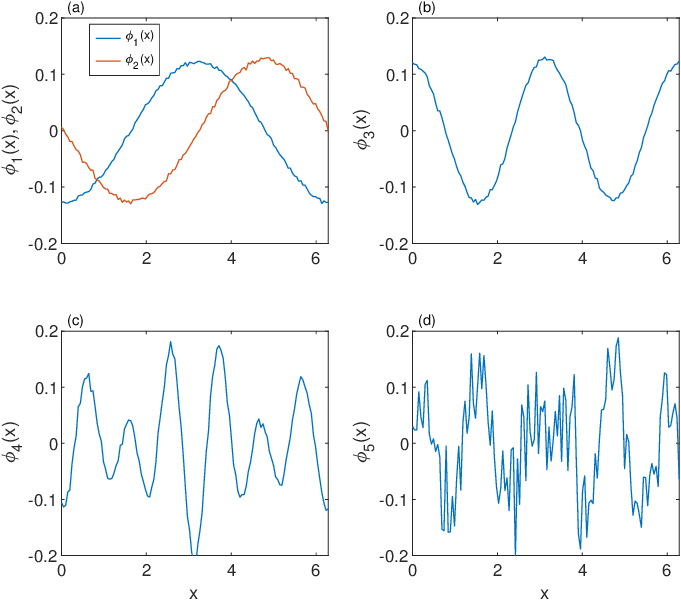}
    \caption{Extracted CCD modes when the observable is sampled using an
	under-resolved frequency $f_s^p=10/2\pi$ while $N=10^4$, $Q=128$ and
	$f_s^u=128/2\pi$. With this low sampling rate, modes 4 and 5 cannot be
    captured accurately.} 
    \label{fig:modes_LowShiftFs}
\end{figure}
In this appendix, we aim to demonstrate the effects of sampling frequency,
duration and multiple observables using the one-dimensional synthetic example
introduced in section~\ref{subsec:1Ddetermin}. We first demonstrate the effects
of varying the sampling frequency of the observable $f_s^p$. In
figure~\ref{fig:modes}, a large sampling frequency of $128/2\pi$ is used. This
is a sufficiently large number considering one only needs to resolve the
approximately maximum frequency of $6/2\pi$ required by the fifth term in
(\ref{equ:artificailFlow}). According to Nyquist's theorem, a minimum sampling
frequency of $12/2\pi$ is needed. To demonstrate the validity of this
requirement, we first perform CCD using $f_s^p=12/2\pi$, resulting in extracted
modes virtually the same as those in figure~\ref{fig:modes}. Subsequently, we
use an under-resolved sampling frequency of $10/2\pi$ while are other parameters
remain unchanged. The resulting modes are shown in
figure~\ref{fig:modes_LowShiftFs}. Clearly, although the low-frequency modes
shown in figure~\ref{fig:modes_LowShiftFs}(a) and (b) can still be correctly
captured, the high-frequency modes shown in figure~\ref{fig:modes_LowShiftFs}(c)
and (d) start to differ from their correct forms. This is because the
under-sampling causes aliasing effects in the decomposition, leading to
incorrect modes $4$ and $5$.

\begin{figure}
    \centering
    \includegraphics[width=0.87\textwidth]{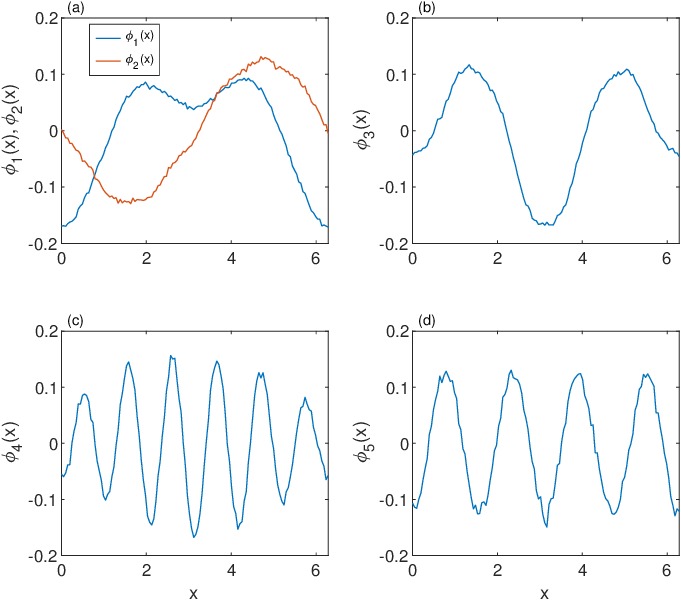}
    \caption{Extracted CCD modes when the observable is sampled for a short
    duration, i.e. $Q=64$, while $N=10^4$ and $f_s^p=f_s^u=128/2\pi$. With this
    short sampling duration, modes 1 and 2 cannot be resolved correctly.} 
    \label{fig:modes_lowShiftDuration}
\end{figure}

We then demonstrate the effects of varying the sampling duration of the
observable. In figure~\ref{fig:modes}, $Q$ is taken as $128$ as this is the
minimum value to use in order to resolve the flow structures given in
(\ref{equ:artificailFlow}), resulting in a frequency resolution of $f_s/Q =
1/2\pi$. If a smaller $Q$ such as $64$ is used, one would expect the failure of
resolving the low-frequency structures. This is precisely the case, as shown in
figure~\ref{fig:modes_lowShiftDuration}, where $Q=64$ while all other parameters
remain the same. As shown in figure~\ref{fig:modes_lowShiftDuration}(a), the
mode at the lowest frequency of $1/2\pi$ is extracted incorrectly. In addition,
the next mode at the frequency of $1/\pi$ seems resolved incorrectly as well
(see figure~\ref{fig:modes_lowShiftDuration}(b)). In fact, the modes shown in
figure~\ref{fig:modes_lowShiftDuration}(a) and (b) appear to have somehow mixed.
This is expected, since the frequency resolution is only $1/\pi$, but the two
modes differ from each other only by $1/2\pi$. On the other hand, modes shown in
figure~\ref{fig:modes_lowShiftDuration}(c) and (d) are characterised by
frequencies of $3/\pi$ and $2/\pi$ respectively, and therefore have been
resolved correctly. Figure~\ref{fig:modes_lowShiftDuration} clearly shows that
the number of shifted rows $Q$ determines the frequency resolution in a similar
manner to that in DFT.

\begin{figure}
    \centering
    \includegraphics[width=0.87\textwidth]{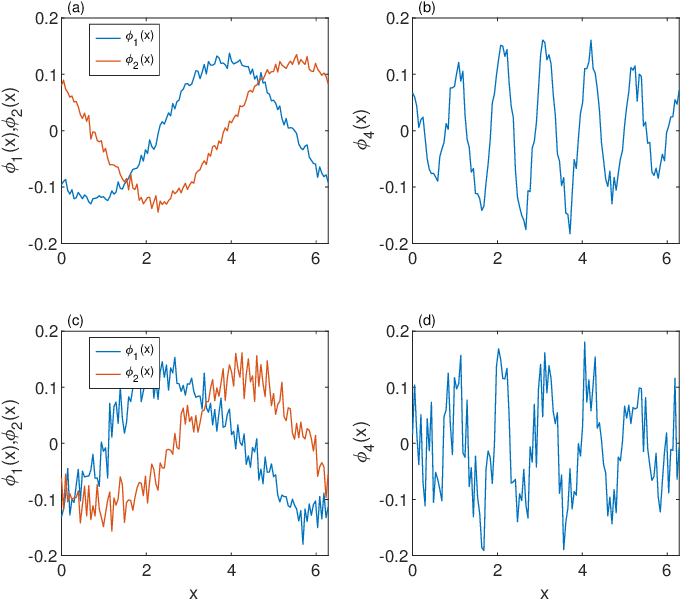}
    \caption{The extracted CCD modes when $N=1000$ (a-b) and $N=100$ (c-d) while
    other parameters remain the same as those in figure~\ref{fig:modes}.
    Convergence increasingly deteriorates as $N$ decreases.} 
    \label{fig:modes_low}
\end{figure}

We are now in a position to illustrate the effects of sampling frequency and
duration of the flow $u$. Figure~\ref{fig:modes} shows that the extracted modes
are subject to small random noise. This noise decreases rapidly as $N$
increases. As mentioned in section~\ref{subsec:resolution}, this is because the
sampling rate and duration of the flow are to ensure the convergence of the
correlation tensor (they do not affect the frequency limit and resolution). In
figures~\ref{fig:toy} and \ref{fig:modes}, $N$ is taken to be $10000$. This is a
large number because we deliberately chose an SNR that is as low as $10^{-4}$. A
small $N$ can also be used at the expense of augmented noise in the resolved
modes. For example, figure~\ref{fig:modes_low} shows the extracted CCD modes
when $N=1000$ (a-b) and $N = 100$ (c-d), respectively. Only modes $1, 2$ and $4$
are shown for brevity. Clearly, we see that using a duration of $N=1000$ results
in CCD modes that converge less well but are still unambiguously identified.
When $N$ reduces to $100$, the resolved CCD modes are further corrupted by the
random noise, nevertheless, the structures of the modes can still be recognized.
Note that $N$ here denotes the number of cycles of the sampled flow. At the
lowest frequency of $100\SI{Hz}$ widely used in the \add{fluid mechanics}
literature, $N=100$ yields a sampling duration of $1\SI{s}$. In the experiments,
a record of $100\SI{s}$ can be easily managed. 

\begin{figure}
    \centering
    \includegraphics[width=0.87\textwidth]{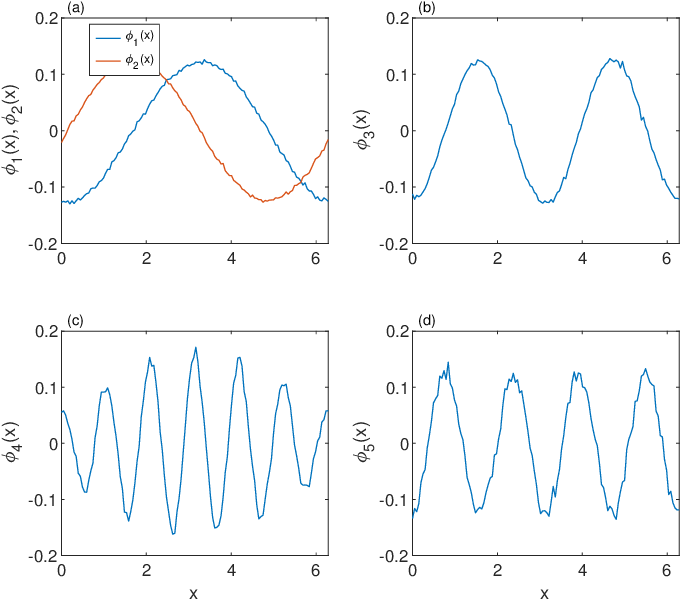}
    \caption{The extracted CCD modes using a low flow sampling frequency of
    $f_s^u=16/2\pi$ while $N=8\times 10^4$, $Q=128$, \add{$f_s^p = 128/2\pi$}.
    Negligible change occurs compared to figure~\ref{fig:modes} because the same
    number of flow snapshots are used for the temporal average.} 
    \label{fig:modes_low_uFs}
\end{figure}
Figure~\ref{fig:modes_low} shows that the CCD modes are corrupted significantly
by random noise at an $\mathrm{SNR}<10^{-4}$ when $N=100$. However, when the
random noise is only two orders of magnitude more energetic, $N=100$ yields
sufficiently well-resolved CCD modes. In general, we find that to obtain the
same level of convergence, $N$ scales roughly as 1/SNR. Conversely, if longer
samples are readily available, CCD can robustly extract the
observable-correlated flow events at the same level of convergence at an even
lower SNR. Therefore, CCD is especially suitable for analysing data acquired in
experiments, where the data may be recorded for as long as desired. 


It is worth noting that although the extracted modes in
figure~\ref{fig:modes_low} are subject to stronger noise contamination due to
the insufficient convergence level, they do not suffer from aliasing effects due
to under sampling. The fact that the sampling frequency and duration of the flow
do not affect the frequency limit and resolution can be even more clearly
demonstrated in figure~\ref{fig:modes_low_uFs}. Figure~\ref{fig:modes_low_uFs}
shows the extracted modes when the flow is only sampled at $16/2\pi$. However,
to exclude the effects of insufficient convergence, the total number of flow
snapshots is kept the same. Comparing figures~\ref{fig:modes} and
\ref{fig:modes_low_uFs} reveals that the extracted modes are virtually
identical, demonstrating the independence of the frequency resolution on the
temporal duration and frequency of sampling.
\begin{figure}
    \centering
    \includegraphics[width=0.87\textwidth]{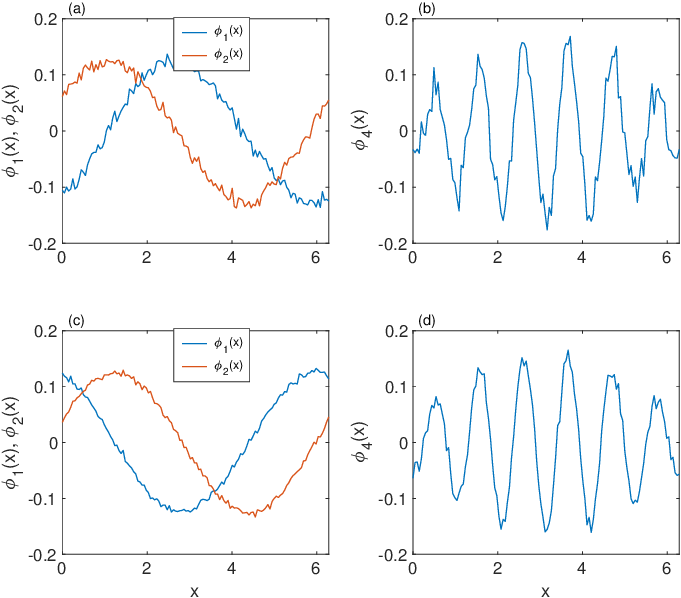}
    \caption{The extracted CCD modes $1$ and $4$ when $1$ (a-b) and $10$ (c-d)
	observables are included. Including multiple observables improves the
    convergences of the resulting CCD modes.} 
    \label{fig:modes_MultiObs}
\end{figure}

We are now in a position to demonstrate the effects of including multiple
observables. The flow field $u$ takes the same form of
(\ref{equ:artificailFlow}). However, more observables need to be defined. To
ensure that the observables are not only similar to (\ref{equ:artificialP}) but
also exhibit variations, we construct up to ten observables $p_i(t)$
($i=1,2,3\ldots {10}$) such that
\begin{equation}
    \begin{aligned}
	p_i(t) = (1+0.2r_{1i})\cos(t-\frac{\pi}{4}) &+
	(1+0.2r_{2i})\sin(2t-\frac{\pi}{3}) + (1+0.2r_{3i}))\cos(4t) \\ &\quad
	\quad \quad \quad \quad + (1+0.2r_{4i})\cos(6t-\frac{\pi}{12}) +
	100r_i(t),
    \end{aligned}
    \label{equ:artificial_multiple_p}
\end{equation}
where $r_{ji}$ ($j=1,2,3,4$) are random numbers between $[-0.5, 0.5]$ and
$r_i(t)$ a random function with a uniform distribution over
$[-0.5, 0.5]$. Note that a very strong random noise (SNR $\sim 10^{-4})$ is also
added in each observable in order to demonstrate the validity of the
decomposition with strongly contaminated observables. 

Following the procedure listed in section~\ref{subsec:multiP}, multiple
observables can be included straightforwardly to form the matrix
$\boldsymbol{P}$. When identical parameters to those in figure~\ref{fig:modes}
are used, the extracted modes using either $1$ or $10$ observables are shown in
figures~\ref{fig:modes_MultiObs}(a,b) and \ref{fig:modes_MultiObs}(c,d),
respectively. Only modes $1, 2$ and $4$ are shown for brevity. Clearly, when the
observables are also strongly contaminated, the extracted modes converge less
satisfactorily, as shown in figure~\ref{fig:modes_MultiObs}(a-b). However, by
including $10$ observables, the quality of the extracted modes improves
significantly. Figure~\ref{fig:modes_MultiObs} shows that including more
observables can indeed improve the convergence of the resulting CCD modes,
particularly when the observables are corrupted by noise. }

\bibliography{cleanRef.bib} 
\bibliographystyle{plainnat}

\end{document}